\documentclass[twocolumn,pra,aps,superscriptaddress, showpacs]{revtex4-2}
\usepackage{amsmath}
\usepackage{amssymb}
\usepackage{amsfonts}
\usepackage{amssymb}
\usepackage[dvips]{graphicx}
\usepackage{subfigure} 
\usepackage{dcolumn}
\usepackage{txfonts}
\usepackage{bm}
\usepackage{makeidx}
\usepackage{color}
\usepackage{mathtools}
\usepackage{threeparttable}
\usepackage[colorlinks,linkcolor=blue,anchorcolor=blue,citecolor=blue,urlcolor=blue]{hyperref}

\begin{document}

\title{Enhanced and tunable electric dipole-dipole interactions near a planar metal film}
\author{Lei-Ming Zhou}
\address{Key Laboratory of Quantum Information, University of Science and
Technology of China, Hefei 230026, China}
\address{Beijing Computational Science Research Center, Beijing 100193, China}
\author{Pei-Jun Yao}
\email{yap@ustc.edu.cn}
\address{University of Science and Technology of China, Hefei 230026, China}

\author{Nan Zhao}
\address{Beijing Computational Science Research Center, Beijing 100193, China}
\author{Fang-Wen Sun}
\email{fwsun@ustc.edu.cn}
\address{Key Laboratory of Quantum Information, University of Science and
Technology of China, Hefei 230026, China}

\date{\textcolor{blue}{July 12, 2017}}
\begin{abstract}
We investigate the enhanced electric dipole-dipole interaction by surface plasmon
polaritons (SPPs) supported by the planar metal film waveguide. By taking two nitrogen-vacancy (NV) center
electric-dipoles in diamond as an example, both the coupling strength and collective relaxation of two
dipoles are studied with numerical Green Function method. Compared to the two-dipole coupling on planar surface, metal film provides stronger and tunable coupling coefficients. Enhancement of the interaction between coupled NV center dipoles could have applications in both quantum information and energy
transfer investigation. Our investigation provides a systematical result for experimental applications
based on dipole-dipole interaction mediated with SPPs on planar metal film.
\end{abstract}

\maketitle

\section{Introduction}
\label{sec:Introduction}

Interaction between photon and atom is one of the most important topics in
modern physics and applications. Mediated by photons, individual atoms can
interact with each other, which has been a kernel in fundamental physics,
ranging from super-radiance \cite{DickePR1954,GrossPR1982,scheibnerNPhysics2007superradiance} to quantum
entanglement \cite{BernienNature2013heralded}. Investigation on interacting atoms or molecules is also
essential for the understanding of energy relaxation and high-efficiency
energy transfer \cite{BouchetPRL2016Long,AndrewScience2004EnergyTansfer,DungPRA2002Intermolecular,MarocicoPRA2011Effect} in
chemical and biological processes \cite%
{YangJCP2010,BennistonMT2008,WasielewskiCR1992}. In addition to natural
atoms and molecules, artificial atoms \cite%
{SipahigilPRL2014,GruberScience1997} with excellent tailorability and
controllability could provide opportunities for studying the light-matter
interaction and for developing devices with novel functions. Among these
artificial atoms, nitrogen vacancy (NV) center in diamond is a very
typical example. NV centers can be generated in diamond crystal with
controllable distance \cite{YamamotoPRB2013}. Accompanying with outstanding
features, such as long decoherence time, stable photon emission, and mature characterization and manipulation techniques by microwave and laser \cite%
{JelezkoPSSA2006,DohertyPR2014}, NV centers can be well applied in quantum
information techniques and biology sensing. Moreover, with the recently
developed super-resolution microscopy techniques \cite{Rittweger2009sted,Cui2013PRL,Chen2014LSA,PfenderPNAS2014SingleSpin}, the quantum state in NV center can be imaged and controlled at nanoscale. Therefore, the coupled NV centers promise an
excellent experimental platform for the study of the photon-mediated dipole-dipole interaction process not only for quantum information techniques, but also
for energy transfer investigations.

In the photon-mediated dipole-dipole interaction, the coupling strength is a key parameter for qubit manipulation in quantum
information techniques and for transfer efficiency in energy transfer
application. In principle, in order to enhance the coupling strength in the dipole-dipole interaction, the photon-atom interaction should be
increased primarily. One of the effective methods to increase the
photon-atom interaction is to couple the atom with surface plasmon polariton
(SPP) on a metal surface \cite{ford1984electromagnetic,GonzalezPRLentanglement,ZhouOL2011Optics} with high density of states and tight spatial
confinement. Recently, researchers have demonstrated that SPP can assist
energy transfer between two ensembles of fluorescent molecules on thin metal
film and the enhancement factor of energy transfer efficiency can be $30$
\cite{BouchetPRL2016Long}. Also, coupling with SPPs on metal surface will
enhance the interaction between two NV centers. Compared to metal nano-wires
\cite{Dzsotjan2011PRB,GonzalezPRLentanglement}, planar metal surface is easy
to fabricate and has potential for scalable applications, based on that
array of NV centers can be generated by ion implantation \cite%
{MeijerAPL2005Generation,YamamotoPRB2013,WangPRB2015High-sensitivity} near
diamond surface with developing techniques \cite%
{OhnoAPL2012Engineering,WangNanoscale2016coherence,ClaireNL2016patterned}.

In this paper, we focused on the negatively charged NV centers in diamond near
various surfaces and systematically studied the coupling strength of coupled
NV centers via their electric dipoles. In detail, the collective relaxation
and coherent coupling of two NV centers near to air surface, metal
surface and metal film are studied numerically with Green Function
(GF) method. It is found that the coupling strength between two NV centers
can be enhanced by the metal surface because of the SPPs. Also, the SPPs on
both sides of the metal film can mediate the coupling of the two dipoles
and thus further enhance the total coupling. The thickness of the metal film can
provide a way to tune the coupling strength. Our investigation will present a systematical result for
the photon-mediated atoms interaction on metal surface. It also provides detailed information for experimental
applications based on coupling NV centers mediated with SPPs on metal film.

This paper is organized as follows. Section~\ref{sec:model-method} describes
the physical model of dipoles of NV centers, and the method we use to
calculate the collective relaxation and coherent coupling of two dipoles. In
Sec.~\ref{sec:dielectric-surface}, we discuss two dipoles near to
air surface. The coherent coupling and collective relaxation of two
NV centers near to diamond and air interface are calculated. The lifetime of
single dipole near to surface is also studied. In Sec.~\ref{sec:
metal-surface}, we discuss two NV centers near to gold and silver surface,
as in Sec.~\ref{sec:dielectric-surface}. The enhancement of dipole-dipole
coupling is studied. In Sec.~\ref{sec:metal-slab}, two NV centers near to
gold film are investigated and the cases of symmetrical and
asymmetrical planar waveguide are both discussed. Finally, Sec.~\ref%
{sec:Conclusion} is the conclusion.

\section{Physical model and method}
\label{sec:model-method}

\begin{figure}[tbp]
\centering
\includegraphics[width=7cm]{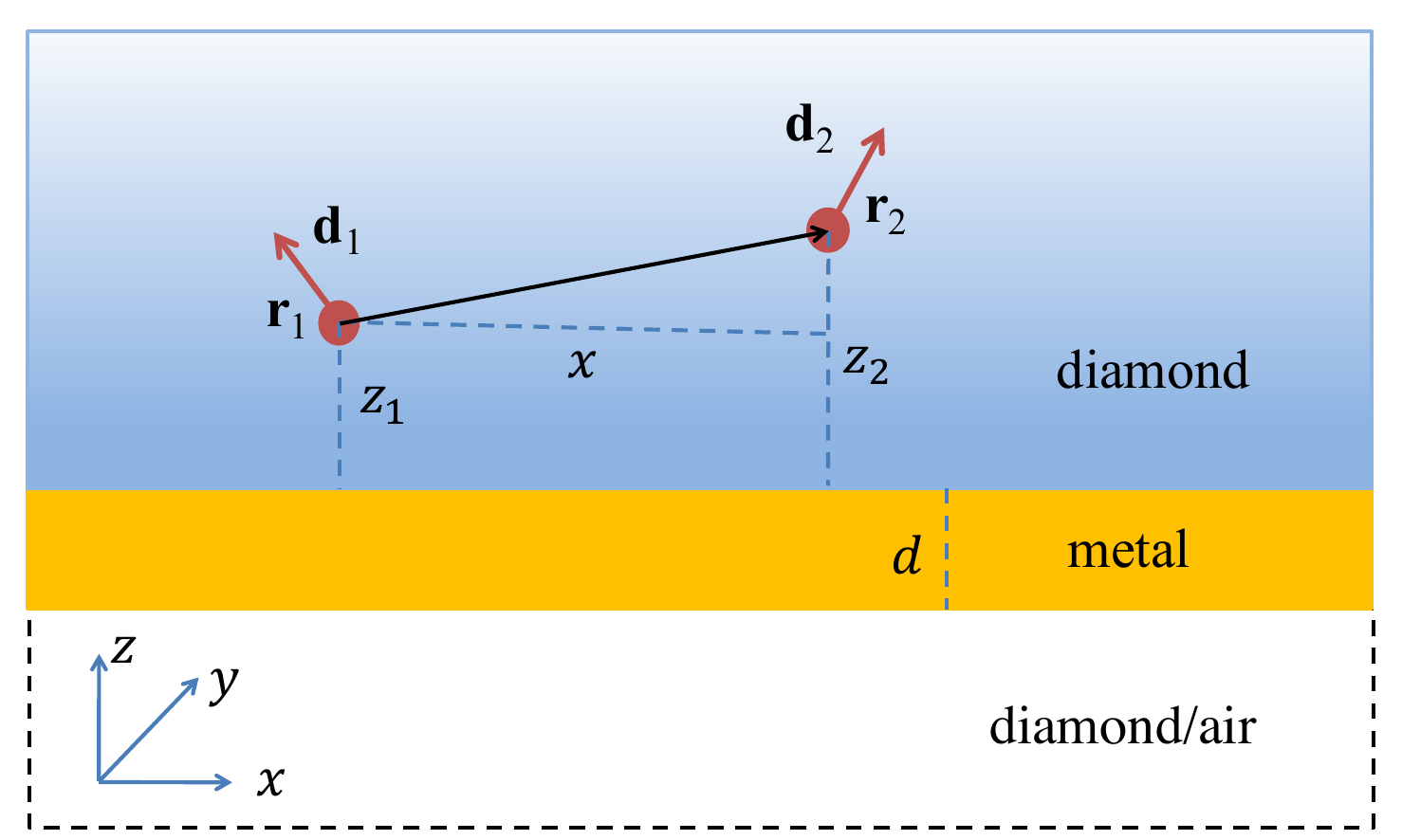}
\caption{Schematic of two dipoles (e.g., dipoles of NV centers) near metal
surface. Two NV centers in diamond crystal near surface with electric
dipoles $\mathbf{d}_{1}$ and $\mathbf{d}_{2}$ respectively are close to each
other (denoted by $NV1$ and $NV2$). The thickness of the metal film is $d$. Under the metal film, it is
air or diamond.}
\label{fig:geometry}
\end{figure}

Two NV centers close to each other in diamond crystal are considered here,
as shown in Fig.~\ref{fig:geometry}. This is usually the case of implanted
NV centers in experiment \cite{YamamotoPRB2013}. One
of the two NV centers is assumed to be located at $\mathbf{r}_{1}=(0,0,z_{1})
$ (denoted as $NV1$), and the other (denoted as $NV2$) is located at $%
\mathbf{r}_{2}=(x,0,z_{2})$ with a displacement $\mathbf{r}=r(\sin \theta
\cos \phi ,\sin \theta \sin \phi ,\cos \theta )$ with respect to $NV1$,
where $r$ is the distance between the two NV centers, and $\theta $ and $%
\phi $ are the polar angle and azimuthal angle of the displacement vector,
respectively. For simplicity, $\phi =0$ is chosen here. NV center is a
multi-level artificial atom and there are linearly polarization dipoles and
circularly polarization dipoles with different directions \cite{MazeNJP2011}%
. The energy levels and allowed optical transitions of NV centers have been
carefully discussed \cite{MansonPRB2006,MazeNJP2011}.

Here, GF method from the electrodynamics is applied to study the
coupling of NV centers. The collective relaxation and coherent coupling
strength of two dipoles are described as \cite{Dzsotjan2011PRB,GonzalezPRLentanglement}:
\begin{align}
\Gamma ^{(12)}& =\frac{\omega _{A}^{2}}{c^{2}\hbar \varepsilon _{0}}\mathrm{%
Im}[\mathbf{d}_{1}\cdot \overleftrightarrow{G}(\mathbf{r}_{1},\mathbf{r}%
_{2},\omega _{A})\cdot \mathbf{d}_{2}]\text{,} \\
\Omega ^{(12)}& =\frac{1}{{\pi }\hbar \varepsilon _{0}}\mathbf{P}%
_{c}\int_{0}^{\infty }\frac{\omega ^{2}}{c^{2}}\frac{\mathrm{Im}[\mathbf{d}%
_{1}\cdot \overleftrightarrow{G}(\mathbf{r}_{1},\mathbf{r}_{2},\omega
_{A})\cdot \mathbf{d}_{2}]}{\omega -\omega _{A}}\mathrm{d}\omega \text{,}
\end{align}%
where $\overleftrightarrow{G}(\mathbf{r}_{1},\mathbf{r}_{2},\omega _{A})$ is
the dyadic Green Function tensor of the electromagnetic field, $\omega _{A}$
is the transition frequency of the dipole, $c$ is the light velocity in
vacuum and $\mathbf{P}_{c}$ denotes the Cauchy value integral. It is noticed
that we use $\Gamma $ to denote the spontaneous damping rate of state
amplitude, which adds a factor $1/2$ compared to that of Ref.~\cite%
{Dzsotjan2011PRB}. It is supposed that two dipoles $\mathbf{d}_{1}$ and $%
\mathbf{d}_{2}$ are with same transition frequency.

The Green function,
$\overleftrightarrow{G}$, gives a full description of the field and the
coupling of dipoles. However, solving the full $\overleftrightarrow{G}$ for
certain structure is usually difficult and numerical solution of the
electromagnetic field equation is needed. Usually, the full $%
\overleftrightarrow{G}$ can be written as
\begin{equation}
\overleftrightarrow{G}=\overleftrightarrow{G_{0}}+\overleftrightarrow{G_{s}},
\label{eq:Gcompose}
\end{equation}%
where $\overleftrightarrow{G_{0}}$ is the free GF of dipole in isotropic
homogenous medium and $\overleftrightarrow{G_{s}}$ is the scattering GF by the structure.

Single dipole couples with the electromagnetic field modes and decays from
the excited state with spontaneous emission rate. When $%
\mathbf{r}_{1}=\mathbf{r}_{2}$, the collective relaxation reduced to the
spontaneous damping rate $\Gamma =1/(2\tau )$, where $\tau $ is the lifetime
of dipole and can be measured by experiment. When the dipole is far away
from the interface, the scattering GF can be eliminated and the spontaneous
damping rate approaches
\begin{equation}
\Gamma _{0}=\frac{\omega _{A}^{2}}{c^{2}\hbar \varepsilon _{0}}\mathrm{Im}[%
\mathbf{d}_{1}\cdot \overleftrightarrow{G_{0}}(\mathbf{r}_{1},\mathbf{r}%
_{1},\omega _{A})\cdot \mathbf{d}_{1}]=\frac{n\omega _{A}^{3}d^{2}}{6{%
\pi }\varepsilon _{0}\hbar c^{3}}\text{,}
\end{equation}%
which is spontaneous damping coefficient in isotropic homogeneous medium.
For dipoles of NV center in diamond, $n=2.418$ is the refractive index. The spontaneous
emission lifetime is $\tau _{0}=1/(2\Gamma _{0})=13.2$ $\mathrm{ns}$
typically.

Also, it is noticed that
\begin{subequations}
\begin{eqnarray}
\Gamma ^{(12)} &=&\Gamma _{0}^{(12)}+\Gamma _{s}^{(12)}\text{,}
\label{eq:couplingDecomposeG} \\
\Omega ^{(12)} &=&\Omega _{0}^{(12)}+\Omega _{s}^{(12)}\text{,}
\label{eq:couplingDecomposeO}
\end{eqnarray}%
according to Eq.~(\ref{eq:Gcompose}). $\Gamma
_{s}^{(12)}$ and $\Omega _{s}^{(12)}$ with subscripts $s$ denote the coupling and relaxation
coefficients induced by the scattering GF. When $r$ is not very small, the
 relaxation rate $\Gamma _{s}^{(12)}$ and coherent coupling strength $\Omega _{s}^{(12)}$ can always be written as \cite{Dzsotjan2011PRB}

\end{subequations}
\begin{subequations}
\begin{align}
\Gamma _{s}^{(12)}& =\frac{\omega _{A}^{2}}{c^{2}\hbar \varepsilon _{0}}%
\mathrm{Im}[\mathbf{d}_{1}\cdot \overleftrightarrow{G_{s}}(\mathbf{r}_{1},%
\mathbf{r}_{2},\omega _{A})\cdot \mathbf{d}_{2}]\text{,}
\label{eq:GFrelaxScat} \\
\Omega _{s}^{(12)}& =\frac{\omega _{A}^{2}}{c^{2}\hbar \varepsilon _{0}}%
\mathrm{Re}[\mathbf{d}_{1}\cdot \overleftrightarrow{G_{s}}(\mathbf{r}_{1},%
\mathbf{r}_{2},\omega _{A})\cdot \mathbf{d}_{2}]\text{.}
\label{eq:GFcoupScat}
\end{align}

Since the free GF $\overleftrightarrow{G}_{0}$ has an explicit expression (see Appendix~\ref{sec:Green-function}), the coherent coupling and
collective relaxation contributed by $\overleftrightarrow{G}_{0}$ can be
obtained easily, with similar expressions in Eq.~(\ref{eq:GFrelaxScat}) and
Eq.~(\ref{eq:GFcoupScat}). In the following, we will focus on the scattering GF in Eq.~(\ref{eq:GFrelaxScat}) and Eq.~(\ref{eq:GFcoupScat}).
There are several numerical methods to calculate the scattering GF of the
system. For stratified medium, the method based on the Fresnel Law of
electromagnetic field and the Wely Identity is an effective one \cite{PaulusPRE2000,novotny2012principles} (see Appendix~\ref{sec:stratifiedGF}). It is noted that the theory here is valid for arbitrary depth and directions of dipoles, as the equations formulated in this work. For numerical calculation, we assume that two dipoles are with the same depth and consider only some typical cases in real experiment, where the two dipoles are parallel.

\section{NV centers near to air surface}
\label{sec:dielectric-surface}

\begin{figure}[tbp]
\centering
\includegraphics[width=8.5cm]{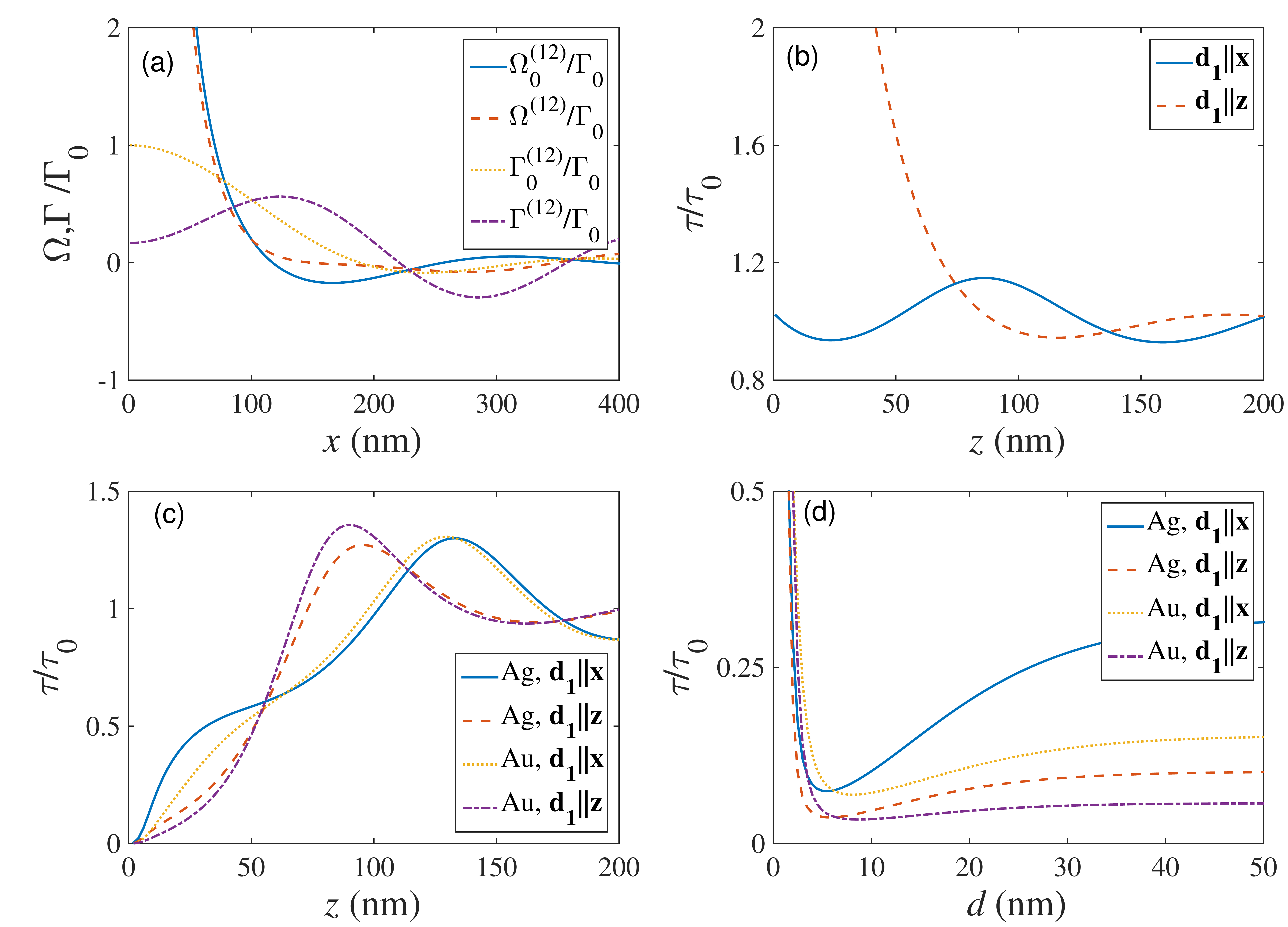}
\caption{(a) The coherent coupling $\Omega^{(12)}$ and collective relaxation
$\Gamma^{(12)}$ for two NV center dipoles in diamond near to air surface,
with different displacement along $x$. Two dipoles are both perpendicular to
surface, e.g., $\mathbf{d}_{1}||\mathbf{z},\mathbf{d}_{2}||\mathbf{z}$, and
are both with depth $16$~$\mathrm{nm}$, e.g., $z_{1}=z_{2}=16$~$\mathrm{nm}$. (b) Lifetime of single
NV center dipole $\mathbf{d}_{1}$ in diamond near to air surface, with
different depth $z$. Both the cases of $\mathbf{d}_{1}||\mathbf{x}$ and $%
\mathbf{d}_{1}||\mathbf{z}$ are shown. It is noted that the lifetime of the $%
\mathbf{d}_{1}||\mathbf{z}$ dipole is not divergent but approaches $16.1$ when
distance $z$ approaches $0$. Similar phenomenon of lengthening the lifetime
here can also be seen in Ref.~\protect\cite{RuppinJCP2004}. (c) The lifetime
of single dipole $\mathbf{d}_{1}$ in diamond near to metal surface, with
different depth $z$. Here, metal (gold or silver) has filled the
semi-infinite space below the $z=0$ interface. (d) The lifetime of single
dipole $\mathbf{d}_{1}$ in diamond near to metal film surface, with
different film thickness $d$. The dipoles are assumed to be located with
depth $16$~$\mathrm{nm}$ in diamond ($z_{1}=16$~$\mathrm{nm}$). In Fig.(c) and (d), both the cases
of dipoles parallel ($\mathbf{d}_{1}||\mathbf{x}$) and perpendicular ($%
\mathbf{d}_{1}||\mathbf{z}$) to surface are shown.
\label{fig:Lifetime-Coupling}
}
\end{figure}

NV centers in diamond are usually generated by ion implantation, with depth
between several nanometers to tens of nanometers. For example, under the $20$~$\mathrm{KeV}$ energy ion-implantation, the depth of NV centers is
distributed among $8\sim 20$~$\mathrm{nm}$ with a maximum at about $16$~$\mathrm{nm}$ \cite{YamamotoPRB2013}. The coupling of two dipoles near to
diamond and air interface is calculated in this section.

The two dipoles can
have different directions. For simplicity, two dipoles both along $z$ axis
are considered first. The collective relaxation and coherent coupling coefficients induced by scattering GF can be expressed as

\end{subequations}
\begin{subequations}
\begin{eqnarray}
\Gamma _{s}^{(12)} &=&\Gamma _{0}\frac{6{\pi }}{k_1}\mathrm{Im}%
[G_{s,zz}(\mathbf{r}_{1},\mathbf{r}_{2},\omega _{A})]  \notag \\
&=&\Gamma _{0}\mathrm{Re}[\int_{0}^{\infty }\mathrm{d}\xi J_{0}(k_{1}x\xi)%
\frac{\xi^{3}r^{p}\mathrm{e}^{\mathrm{i}k_{1}\sqrt{1-\xi^{2}}(z_{1}+z_{2})}}{%
\sqrt{1-\xi^{2}}}]\text{,} \\
\Omega _{s}^{(12)} &=&\Gamma _{0}\frac{6{\pi }}{k_1}\mathrm{Re}%
[G_{s,zz}(\mathbf{r}_{1},\mathbf{r}_{2},\omega _{A})]  \notag \\
&=&\Gamma _{0}\mathrm{Im}[\int_{0}^{\infty }\mathrm{d}\xi J_{0}(k_{1}x\xi)%
\frac{\xi^{3}r^{p}\mathrm{e}^{\mathrm{i}k_{1}\sqrt{1-\xi^{2}}(z_{1}+z_{2})}}{%
\sqrt{1-\xi^{2}}}]\text{.}
\end{eqnarray}%
They are single variable integrals and can be calculated numerically. Among them, $%
k_1$ is the wavevector in the diamond, $x$ is the horizontal displacement as shown in Fig.~\ref{fig:geometry}, $J_0$ is the Bessel function, $\xi$ is the integration variable. $G_{s,zz}$ is the element of $\overleftrightarrow{G}_{s}$. $r^{s}$ and $r^{p}$ are the Fresnel
reflection coefficients for $s$ and $p$ polarized light, respectively. Here, since both the two dipoles are along $z$ axis, $r^s$ doesn't appear (see Appendix~\ref{sec:stratifiedGF}).

The
total coupling and relaxation coefficients according to Eq.~(\ref%
{eq:couplingDecomposeG}) and Eq.~(\ref{eq:couplingDecomposeO}) are calculated
and shown in Fig.~\ref{fig:Lifetime-Coupling}(a). We can see that, the
collective relaxation coefficient near to surface is very different from
that in diamond crystal. Due to the reflection of the free GF from the
interface, the total GF changes considerably.

The relaxation rate ($\Gamma$) of single NV center with
dipole $\mathbf{d}_{1}$ is also affected by surface and is discussed here. Normalized by $\Gamma
_{0}$, it is

\end{subequations}
\begin{equation}
p=\frac{\Gamma }{\Gamma _{0}}=\frac{\mathrm{Im}[\mathbf{d}_{1}\cdot
\overleftrightarrow{G}\cdot \mathbf{d}_{1}]}{\mathrm{Im}[\mathbf{d}_{1}\cdot
\overleftrightarrow{G_{0}}\cdot \mathbf{d}_{1}]}=1+\frac{6{\pi }}{k_1}%
\mathrm{Im}[\hat{d}_{1,i}G_{s,ij}\hat{d}_{1,j}\}] \text{,}  \label{eq:gammaEnhancep}
\end{equation}%
where $\hat{d}_{1}$ is the normalized direction of dipole $\mathbf{d}_{1}$
and $i,j=x,y,z$ denote the Cartesian components of the dipole $\mathbf{d}_{1}$%
. When the dipole is perpendicular to the interface, i.e., $\mathbf{d}%
_{1}\Vert \mathbf{z}$, only $i=z$ component is remained and $p_{\perp }=1+%
\frac{6{\pi }}{k_1}\mathrm{Im}[G_{s,zz}]$, where $p_{\perp }$ denotes
the relaxation enhancement of dipole perpendicular to the surface.
Similarly, the case of dipole parallel to the surface is computed as $%
p_{\parallel }.$ Both $p_{\perp }$ and $p_{\parallel }$ are numerically
calculated and shown in Fig.~\ref{fig:Lifetime-Coupling}(b). Here, the GF
reflected by the surface interferes with the free GF, so the relaxation rate
(or the lifetime) oscillates with a period of half wavelength of the photons from NV center electric dipole.

For dipoles with angle $\alpha $ with $z$ axis, Eq.~(\ref{eq:gammaEnhancep})
is simplified as
\begin{equation}
p=1+sin^{2}(\alpha )p_{\parallel }+cos^{2}(\alpha )p_{\perp }.
\end{equation}%
There are dipoles of various directions in NV center ensemble, and the lifetime measurement in experiment gets the average lifetime of all kinds of dipoles.

\section{NV centers near to metal surface}
\label{sec: metal-surface}

The metal with negative dielectric coefficient supports SPPs when $%
\varepsilon _{metal}<-\varepsilon _{dielectric}$. Dipoles near to metal surface
can couple to SPPs and have large spontaneous damping rate. Two dipoles near
metal surface can also have stronger coupling through the SPPs compared to
that through the electromagnetic field in free space. Array of NV centers
can be generated with well designed ion-implantation techniques and metal
film can be deposited with controllable thickness. These techniques promise
that array NV centers in diamond can be applied in the study of SPP enhanced electric dipole-dipole interaction for quantum information and energy transfer.

For the interface between dielectric material with dielectric constant $%
\varepsilon _{1}$ and metal material with dielectric constant $\varepsilon
_{2}$, the SPP modes have a wave vector
\begin{equation}
k_{sp}=k_{0}\sqrt{\frac{\varepsilon _{1}\varepsilon _{2}}{\varepsilon
_{1}+\varepsilon _{2}}}=k_{1}\sqrt{\frac{\varepsilon _{2}}{\varepsilon
_{1}+\varepsilon _{2}}}\text{,}
\end{equation}%
where $k_{0}=2{\pi }/\lambda _{0}$ is vacuum wavevector, $k_{1}=2%
{\pi }/\lambda _{1}$ is wave vector in dielectric material and $\lambda _{1}=\lambda _{0}/n$ is
the wavelength in diamond here. For NV
centers in diamond, the zero phonon line (ZPL) is $\lambda _{0}=637.2$~$%
\mathrm{nm}$. The dielectric coefficient of Au is $\varepsilon
_{2}=-10.85+1.27\mathrm{i}$ at this wavelength \cite{PolyanskiyWebBookRefractive}. The imaginary part of the
dielectric coefficient results in both phase change and absorption.
$\varepsilon _{1}=5.847$ is the dielectric coefficient of diamond. So, in
this system the SPP mode has a wavelength $\lambda _{sp}=181.5$~$\mathrm{nm}$.

\begin{figure}[htbp]
\centering
\includegraphics[width=8.5cm]{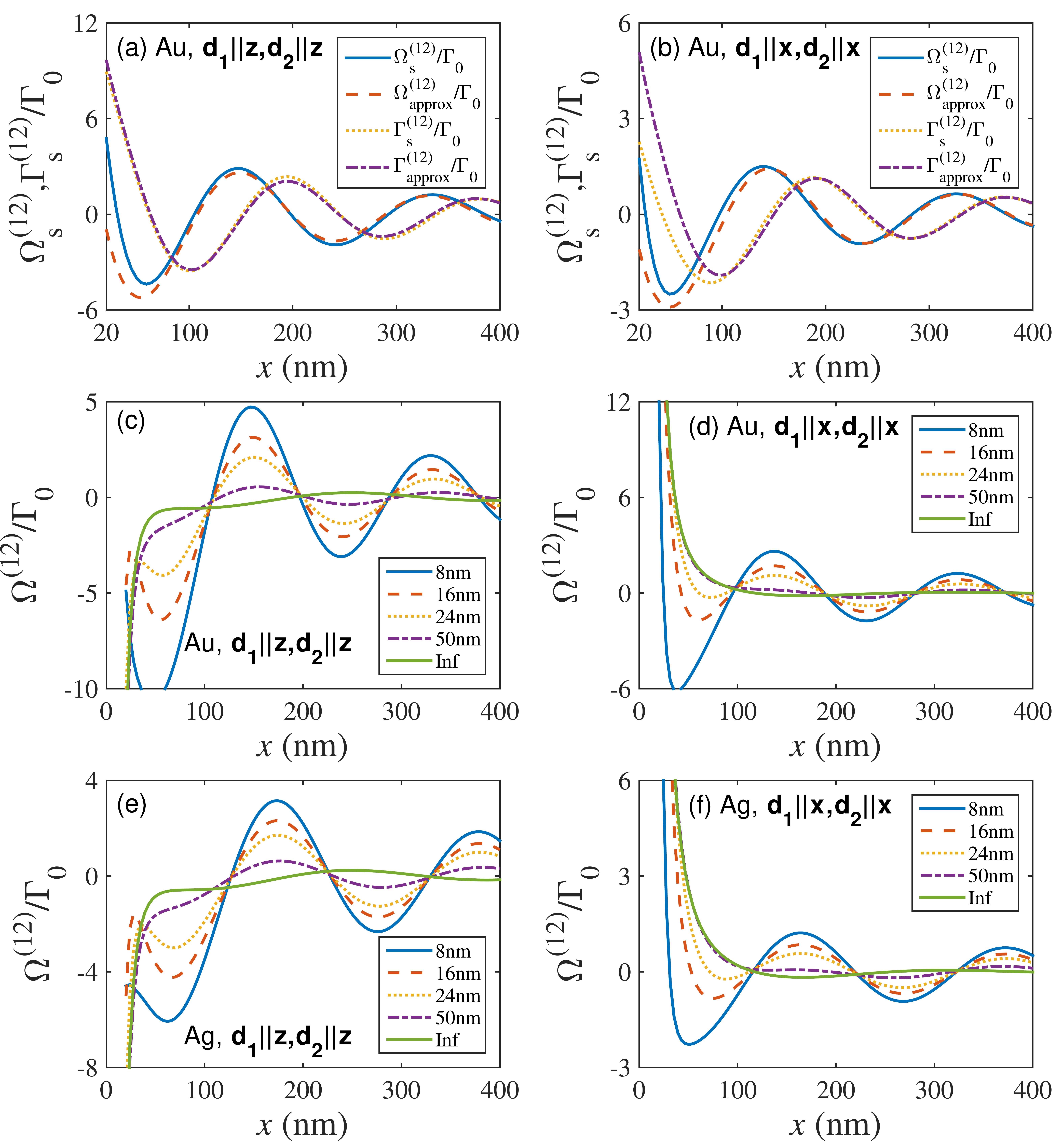}
\caption{(a)(b) The coherent coupling coefficient $\Omega_{s}^{(12)}$ and
collective relaxation $\Gamma_{s}^{(12)}$ for two dipoles in diamond on Au
surface, with different displacement along $x$. Two dipoles are both located
with depth $20$~$\mathrm{nm}$ in diamond ($z_{1}=z_{2}=20$~$\mathrm{nm}$). The $y$ direction
displacement of two dipoles is 0. Theoretic approximations are shown for
comparison. (a) is for two dipoles both
perpendicular to surface, $\mathbf{d}_{1}||\mathbf{z},\mathbf{d}_{2}||%
\mathbf{z}$. (b) is for two dipoles both parallel to surface, $\mathbf{d}%
_{1}||\mathbf{x},\mathbf{d}_{2}||\mathbf{x}$. (c)(d)(e)(f) The coupling
coefficient $\Omega^{(12)}$ for two dipoles in diamond on gold and silver
surface, with different displacement along $x$. Two dipoles both with depth $%
8$~$\mathrm{nm}$, $16$~$\mathrm{nm}$, $24$~$\mathrm{nm}$ and $50$~$\mathrm{nm}$ are shown. The depth $Inf$ is for two dipoles far away from surface, i.e., in
homogeneous diamond material. (c) is for two dipoles both perpendicular
to gold surface, $\mathbf{d}_{1}||\mathbf{z},\mathbf{d}_{2}||\mathbf{z}$.
(d) is for gold surface, $\mathbf{d}_{1}||\mathbf{x},\mathbf{d}_{2}||%
\mathbf{x}$. (e) is for silver surface, $\mathbf{d}_{1}||\mathbf{z},%
\mathbf{d}_{2}||\mathbf{z}$. (f) is for silver surface, $\mathbf{d}_{1}||%
\mathbf{x},\mathbf{d}_{2}||\mathbf{x}$. }
\label{fig:CouplingonMetalDepth}
\end{figure}

The coherent coupling and collective relaxation of two dipoles on metal
surface are shown in Fig.~\ref{fig:CouplingonMetalDepth}. The case of both
dipoles perpendicular to the surface, i.e., $\mathbf{d}_{1}||\mathbf{z},%
\mathbf{d}_{2}||\mathbf{z}$, is shown in Fig.~\ref{fig:CouplingonMetalDepth}%
(a), while Fig.~\ref{fig:CouplingonMetalDepth}(b) is for $\mathbf{d}_{1}||%
\mathbf{x},\mathbf{d}_{2}||\mathbf{x}$. Here, we consider only the
scattering GF induced coupling. In principle, the free space coupling should
be added to get the total coupling. However, it can be neglected when the
dipoles are near to surface and the distance between two dipoles are not
very small. We can see that the coupling is much stronger than that in free space
or on dielectric surface. Since the SPP modes have approximate analytic
solutions \cite{AbajoRMP2007,RotenbergPRL2012}, the approximated expression
of relaxation and coupling can also be expressed as
\begin{eqnarray}
\frac{\Gamma _{s}^{(12)}}{\Gamma _{0}} &=&\frac{6{\pi }}{k_{1}}%
\mathrm{Re}[G_{xx}(\mathbf{r}_{1},\mathbf{r}_{2},\omega _{A})] \\
&\approx &c_{1}\mathrm{Im}[c_{2}\mathrm{ie}^{\mathrm{i}%
(k_{sp}x+k_{sp}^{z}(z_{1}+z_{2})-\frac{{\pi }}{4})}/\sqrt{k_{sp}x}]%
\text{,}  \label{eq:relaxationMetal} \\
\frac{\Omega _{s}^{(12)}}{\Gamma _{0}} &=&\frac{6{\pi }}{k_{1}}%
\mathrm{Im}[G_{xx}(\mathbf{r}_{1},\mathbf{r}_{2},\omega _{A})] \\
&\approx &c_{1}\mathrm{Re}[c_{2}\mathrm{ie}^{\mathrm{i}%
(k_{sp}x+k_{sp}^{z}(z_{1}+z_{2})-\frac{{\pi }}{4})}/\sqrt{k_{sp}x}]%
\text{.}  \label{eq:couplingMetal}
\end{eqnarray}%
The approximation results of Eq.~(\ref{eq:relaxationMetal}) and Eq.~(\ref%
{eq:couplingMetal}) are also plotted in Fig.~\ref{fig:CouplingonMetalDepth}(a)(b)
for two dipoles both with $20$~$\mathrm{nm}$ depth, where $c_{1}$ and $c_{2}$ are complex
coefficients independent of $x$. We can see that it is indeed a good
approximation when the dipole depth is much smaller than the wavelength.

The coupling coefficient will increase when the two dipoles approach the
metal surface. As we can
see in Fig.~\ref{fig:CouplingonMetalDepth}(b)(c), the coupling becomes stronger when the distance $z$ decreases. The
oscillating period of coupling coefficient is the surface plasmon wavelength
$\lambda _{sp}$. The line for infinite distance is the total coupling
coefficient of two dipoles in dielectric material. The case for the
silver surface is also studied. The coupling and relaxation of both dipoles
parallel and perpendicular to surface are plotted in Fig.~\ref%
{fig:CouplingonMetalDepth}(e)(f). The dielectric constant of silver is $%
\varepsilon _{Ag}=-14.69+1.21\mathrm{i}$ \cite{PolyanskiyWebBookRefractive} at the wavelength of the ZPL of NV
center.

The lifetime of a dipole near to metal surface is
shown in Fig.~\ref{fig:Lifetime-Coupling}(c). When the distance away from the metal surface is at the scale of wavelength, the lifetime of a dipole is modulated by the
interference. At this time, the dipoles mainly emit free photons mostly. It is similar to the case
of dielectric material. When the dipole is approaching the metal with
distance much less than $\lambda _{1}/2$, it couples to the SPP modes and the lifetime
decreases very much. When distance $z$ is very small, there is dipole
emission quenching. The dielectric coefficient model of metal used here will
be invalid, and thus the lifetime of dipole can not approach to 0 when $%
z\rightarrow 0$.

\section{NV centers near to metal film}
\label{sec:metal-slab}

When the metal material is a film with thickness much less than wavelength,
the SPP modes on both sides of the film can affect the dipole-dipole
interaction and the lifetime of a single dipole. The lifetime of NV center dipoles on
metal film surface (downside of the film is air) is shown in Fig.~\ref{fig:Lifetime-Coupling}(d). When the thickness of film $d>50$~$\mathrm{nm}$, the lifetime of a
dipole stays the same as that on metal surface. For the case of a dipole
perpendicular ($\mathbf{d}_{1}||\mathbf{z}$) to gold surface, when $10$~$%
\mathrm{nm}<d<50$~$\mathrm{nm}$, the SPP modes on the downside of the film
can couple to the dipole and reduce the lifetime. The lifetime is decreased
to a minimum when $d\simeq 10$~$\mathrm{nm}$.
When the thickness $d<10$~$\mathrm{nm}$, the bounding of light to SPP modes
becomes weaker, and the lifetime increases. Lifetime of dipoles parallel to gold film surface and lifetime of dipoles on silver film surface are also shown. For these cases, the lifetime reduces
to minimum at a smaller thickness (about $5$~$\mathrm{nm}$), and other behaviors keep the
same.

\begin{figure}[t]
\centering
\includegraphics[width=8.5cm]{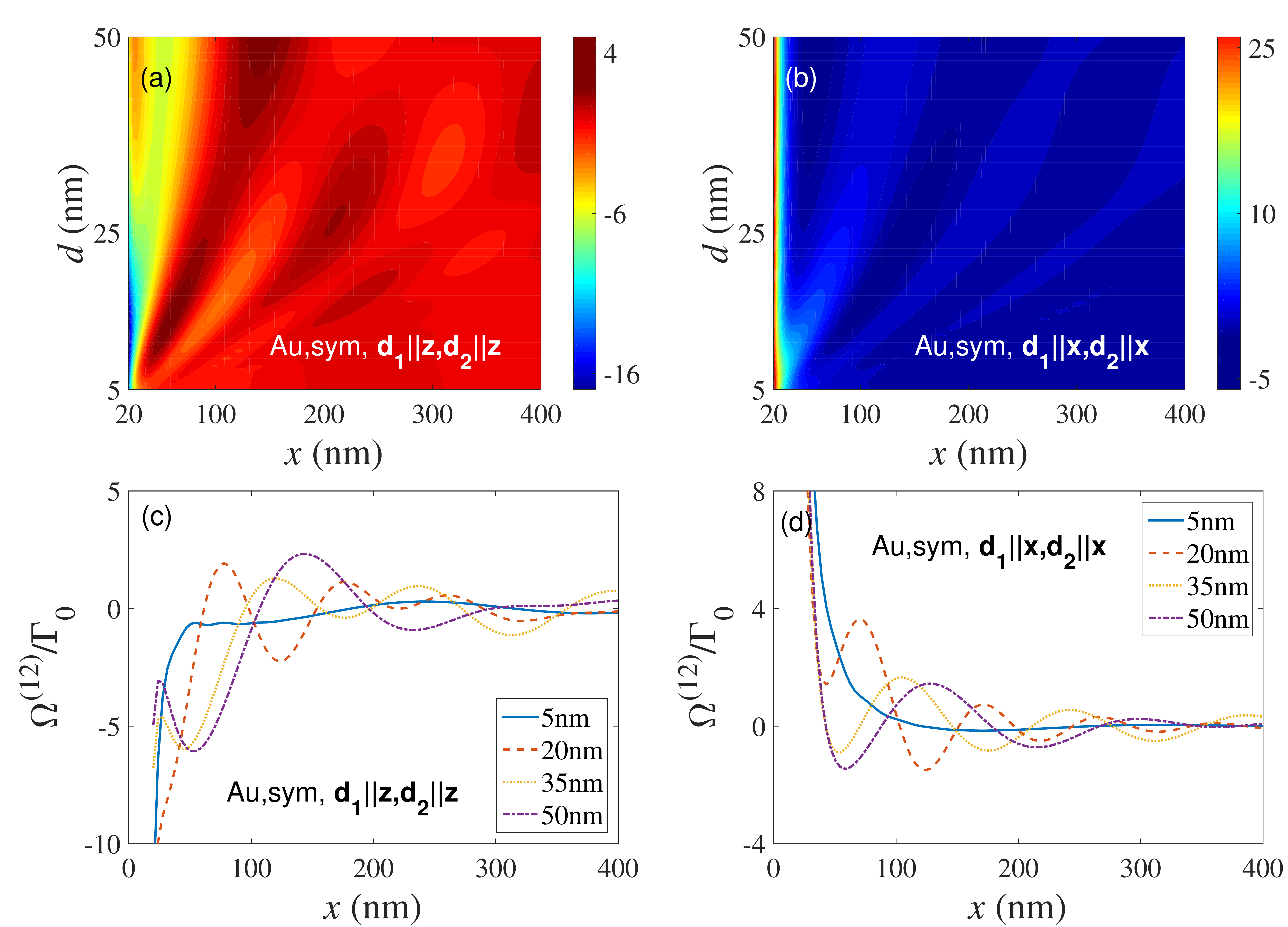}
\caption{The coupling coefficient $\Omega^{(12)}$ for two dipoles in diamond
on symmetrical gold film surface, with different displacement along $x$ and different
film thickness $d$. (a) Two dipoles are both perpendicular to surface, i.e.,
$\mathbf{d}_{1}||\mathbf{z},\mathbf{d}_{2}||\mathbf{z}$. (b) Two dipoles are
both parallel to surface, i.e., $\mathbf{d}_{1}||\mathbf{x},\mathbf{d}_{2}||%
\mathbf{x}$. (c)(d) Line cuttings of (a) and (b), respectively.
The gold film with thickness $d=5$~$\mathrm{nm}$, $20$~$\mathrm{nm}$, $35$~$\mathrm{nm}$ and $50$~$\mathrm{nm}$ are shown. In these figures, two dipoles
are both located at the depth $16$~$\mathrm{nm}$ in diamond ($z_{1}=z_{2}=16$~$\mathrm{nm}$).}
\label{fig:CouplingAuSlab-d}
\end{figure}

\begin{figure}[htbp]
\centering
\includegraphics[width=8.5cm]{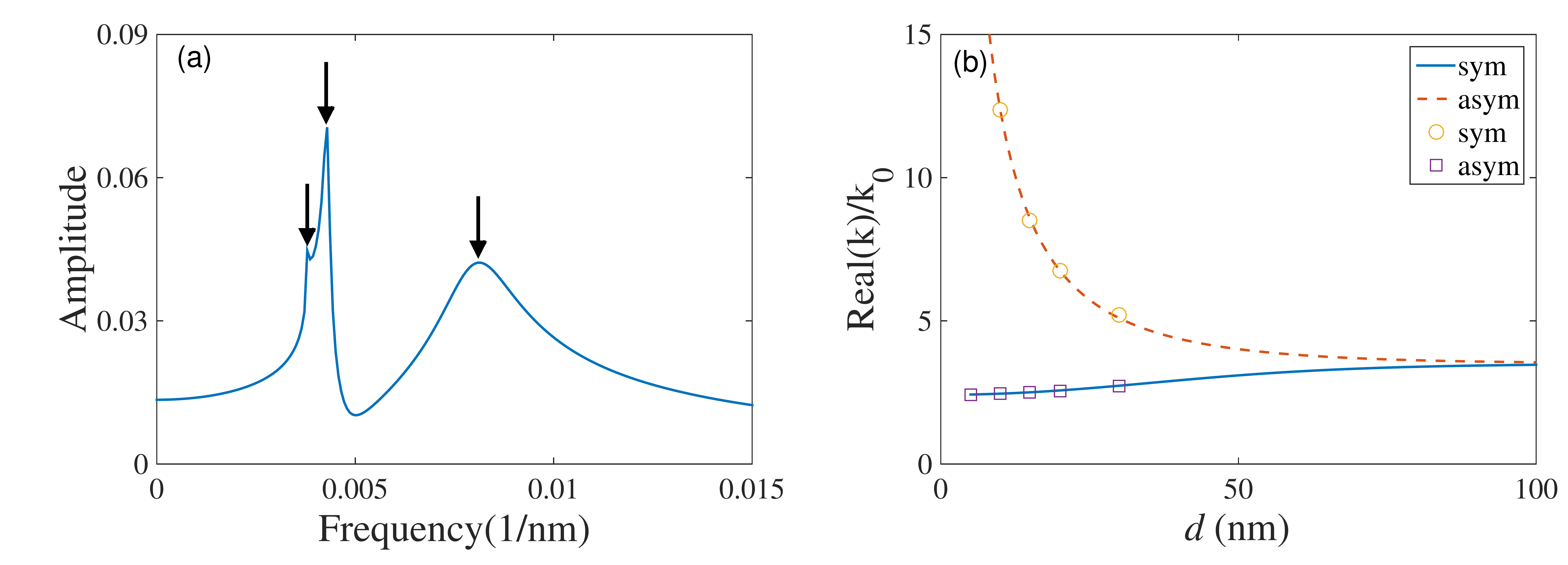}
\caption{The modes mediated in the dipole-dipole interaction and the dispersion
relation of the symmetrical and asymmetrical modes of symmetric gold planar
waveguide. (a) Fourier transform of the coupling coefficient $\Omega^{(12)}$
for two dipoles both with depth $30$~$\mathrm{nm}$ to surface and direction perpendicular to
surface, i.e., $\mathbf{d}_{1}||\mathbf{z},\mathbf{d}_{2}||\mathbf{z}$. (b)
The effective refractive indices of SPP modes supported by the gold film with
different thickness $d$. The lines are the numerical solutions of the
dispersion equation for SPP modes. The symbols are for the modes from (a), with film thickness $%
d=5$~$\mathrm{nm}$, $10$~$\mathrm{nm}$, $15$~$\mathrm{nm}$, $20$~$\mathrm{nm}$ and $30$~$\mathrm{nm}$, respectively.}
\label{fig:dispersion}
\end{figure}

The film thickness dependence of dipole-dipole interaction near to symmetrical metal film is shown in Fig.~\ref%
{fig:CouplingAuSlab-d}(a)(b). And Fig.~\ref{fig:CouplingAuSlab-d}(c)(d) are
some line cuttings of Fig.~\ref{fig:CouplingAuSlab-d}(a)(b) with certain thicknesses. It can be seen
that, with the decrease of the film thickness, the frequency of the SPP
increases.

For symmetrical metal film with same dielectric constant on both sides, the
coupling between two dipoles is induced by free
electromagnetic field modes in diamond, symmetrical SPP mode and
asymmetrical SPP mode of the metal film. To show this, the Fourier transform
of the coupling strength on the symmetrical metal film with thickness $30$~$\mathrm{nm}$
is shown in Fig.~\ref{fig:dispersion}(a). The three maxima corresponds to
three modes at spatial frequency $0.0038/\mathrm{nm}$, $0.0043/\mathrm{nm}$,
and $0.0081/\mathrm{nm}$, with effective refractive indices $2.426$, $%
2.745$, and $5.170$, respectively. It is  analyzed that they are free mode, symmetrical mode and asymmetrical mode respectively.
The cases for film thickness with $d=5$~$\mathrm{nm}$, $10$~$\mathrm{nm}$, $15$~$\mathrm{nm}$, $20$~$\mathrm{nm}$ and $30$~$\mathrm{nm}$ are also
calculated. The symmetrical and asymmetrical modes are shown in Fig.~\ref{fig:dispersion}(b). The solid and dashed
lines in Fig.~\ref{fig:dispersion}(b) are the numerical solutions of the
dispersion equation of symmetrical metal film for SPP modes \cite{BurkePRB1986surface}.
It can be seen that they match very well and verifies our analysis.

\begin{figure}[t]
\centering
\includegraphics[width=8.5cm]{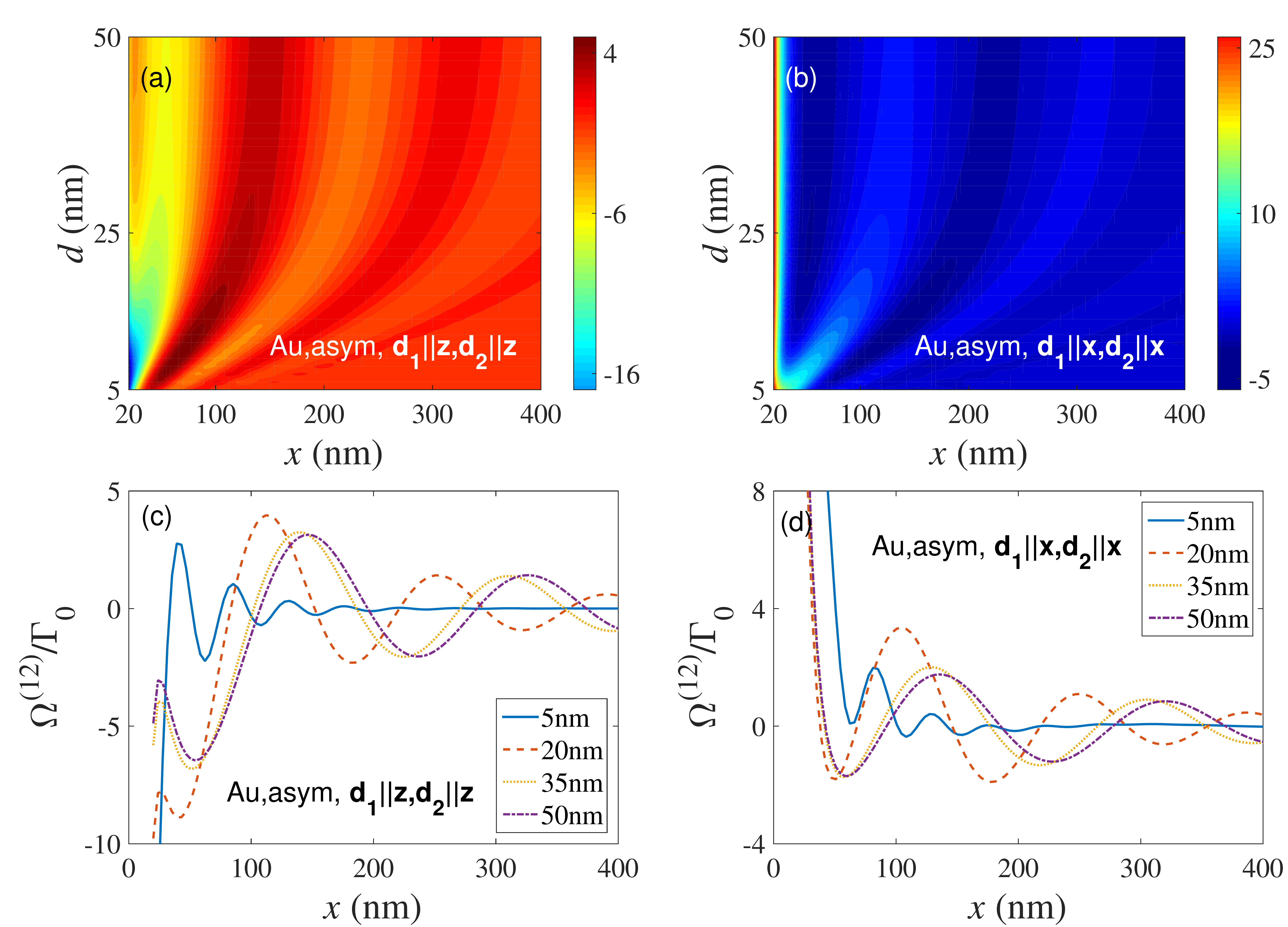}
\caption{The coupling coefficient $\Omega^{(12)}$ for two dipoles in diamond
on asymmetrical gold film surface, with different displacement along $x$ and
different film thickness $d$. The substrate is changed from diamond to air
compared to Fig.~\protect\ref{fig:CouplingAuSlab-d}. Other parameters are the same as Fig.~\protect\ref%
{fig:CouplingAuSlab-d}. }
\label{fig:CouplingAuSlabasym-d}
\end{figure}

The coherent coupling coefficient of two dipoles on asymmetrical metal
planar waveguide is also studied. The result is shown in Fig.~\ref%
{fig:CouplingAuSlabasym-d}. Here, the material below the gold film is changed from
diamond to air, as in Fig.~\ref{fig:geometry}. The phenomenon of coupling on
asymmetrical waveguide is similar with that on symmetrical waveguide.

\begin{figure}[htbp]
\centering
\includegraphics[width=8.5cm]{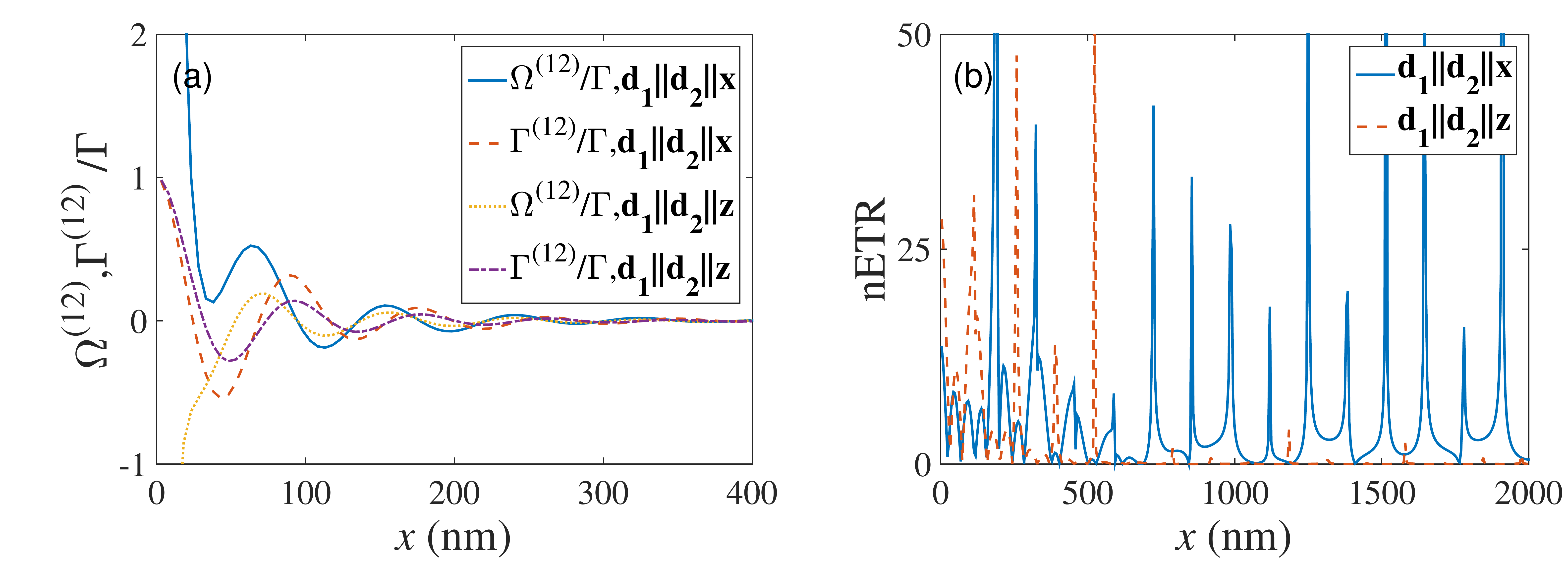}
\caption{(a) $\Omega ^{(12)}$ and $\Gamma ^{12}$ normalized by $\Gamma $ for
two dipoles in diamond on asymmetrical gold film surface, with different
displacement along $x$. The film thickness $d=10$~$\mathrm{nm}$. (b) The normalized
energy transfer rate for two dipoles with directions $\mathbf{d}_{1}||%
\mathbf{d}_{2}||\mathbf{z}$ and $\mathbf{d}_{1}||\mathbf{d}_{2}||\mathbf{x}$%
. }
\label{fig:nETR}
\end{figure}

Fig.~\ref{fig:nETR} shows the enhancement of coherent coupling and the energy transfer for the two dipoles
in diamond on asymmetrical gold film surface. The $\Omega ^{(12)}$ and $%
\Gamma ^{12}$ normalized by $\Gamma $ is shown in Fig.~\ref{fig:nETR}(a). Here, we choose the total decay rate $\Gamma$ of the dipole near the metal film as the normalizing constant, so we can see at which displacement that the coupling exceeds the decay rate. It can be seen that $%
\Omega ^{(12)}/\Gamma $ surpasses $1$ only for distance less than $25$~$\mathrm{nm}$, which means that both the
coupling and relaxation are enhanced by SPPs. 

The normalized energy transfer rate (nETR) between two dipoles are
defined as \cite{MartinNL2010ResonanceEnergy}%
\begin{equation}
nETR=\Gamma ^{(12)}/\Gamma _{0}^{(12)}\text{,}  \label{Eq:nETR}
\end{equation}%
where $\Gamma ^{(12)}$ is the collective
relaxation in the presence of the metal film and $\Gamma _{0}^{(12)}$ is the
collective relaxation in isotropic homogeneous medium (i.e., diamond here).
The results are shown in Fig.~\ref{fig:nETR}(b). It's noted that the dipole directions of NV centers are fixed, which is different from fluorescent molecules. For two fluorescent molecules, the acceptor dipole is induced by the
field from the donor dipole \cite{novotny2012principles,MartinNL2010ResonanceEnergy}. The
nETR should be averaged over various dipole directions \cite{BouchetPRL2016Long}. Thus the nETR defined here for two NV center dipoles is a little different. Here, the two
dipoles are with determined direction and the nETR has sharp points when $%
\Gamma _{0}^{(12)}=0$. The energy transfer rate has been increased by tens of times for both $\mathbf{d}_{1}||\mathbf{d}_{2}||\mathbf{z}$ and $\mathbf{%
d}_{1}||\mathbf{d}_{2}||\mathbf{x}$ for most of distances between the dipoles. Also, nETR for $%
\mathbf{d}_{1}||\mathbf{d}_{2}||\mathbf{x}$ keeps large with distance of $2$~${\mu m}$ and is larger than that for $\mathbf{d}_{1}||\mathbf{d}%
_{2}||\mathbf{z}$. This distance is the same as the distance between
array of NV centers generated by current techniques of ion
implantation \cite{MeijerAPL2005Generation,WangPRB2015High-sensitivity}.

\section{Discussion and Conclusion}
\label{sec:Conclusion}

In conclusion, we have investigated the enhanced electric dipole-dipole interaction near to various surfaces, especially near to the metal planar
film. Starting from the general theory of GF method for
stratified media, the collective relaxation and coherent coupling of two
dipoles of NV centers near to dielectric surface, metal surface and metal
film are calculated. The results verify the enhanced electric-dipole
coupling by SPPs on metal surface and provide a systematical result for experimental applications. Also, it is found that the waveguide
thickness changes the coupling coefficients of two dipoles, through changing of the efficiency and ratio of
the coupling of dipole emissions to different modes of waveguide. It thus
provides a method to tune the coupling of two dipoles. Such a SPP enhanced dipole-dipole interaction in NV center system would be a promising experimental platform for the study of quantum information techniques and energy transfer.

\bigskip
\noindent {\bf{Acknowledgments.}} This work is supported by Strategic Priority Research Program (B)
of the Chinese Academy of Sciences (Grant No. XDB01030200), the National
Natural Science Foundation of China (Grant Nos. 11374032, 11374290, 91536219, 61522508), Anhui Provincial Natural Science Foundation
(Grant Nos. 1508085SMA205, 1408085MKL0), and the Fundamental Research Funds for the Central Universities.

\section{Supplemental Material}
The appendices have been written for the readability, as a concise introduction of the method used in this work. Details can be seen in related references \cite{novotny2012principles,PaulusPRE2000,chew1995waves}.

\subsection{Dyadic Green Function tensor\label{sec:Green-function}}
The dyadic GF $\overleftrightarrow{G_{0}}$ of isotropic homogenous medium
for dipole emission is the solution of
\begin{equation}
\nabla \times \nabla \times \overleftrightarrow{G_{0}}(\mathbf{r}_{1},%
\mathbf{r}_{2},k)-k^{2}\overleftrightarrow{G_{0}}(\mathbf{r}_{1},\mathbf{r}%
_{2},k)=I\delta (\mathbf{r}_{1},\mathbf{r}_{2}) \text{,}
\end{equation}%
and can be explicitly expressed as
\begin{subequations}
\begin{eqnarray}
\overleftrightarrow{G_{0}}(\mathbf{r}_{1},\mathbf{r}_{2},k) &=&[I+\frac{1}{%
k^{2}}\nabla \nabla ]\frac{\mathrm{e}^{\mathrm{i}kr}}{4{\pi}r} \\
&=&\frac{\mathrm{e}^{\mathrm{i}kr}}{4{\pi}r}[(1-\frac{1}{\mathrm{i}kr%
}-\frac{1}{(kr)^{2}})I  \notag \\
&&+(-1+\frac{3}{\mathrm{i}kr}+\frac{3}{(kr)^{2}})\hat{r}\hat{r}]+\frac{%
\delta (r)}{3k^{2}}I \\
&=&\frac{k\mathrm{e}^{\mathrm{i}kr}}{4{\pi}}[(\frac{1}{kr}+\frac{%
\mathrm{i}}{(kr)^{2}}-\frac{1}{(kr)^{3}})I  \notag \\
&&-(\frac{1}{kr}+\frac{3\mathrm{i}}{(kr)^{2}}-\frac{3}{(kr)^{3}})\hat{r}\hat{%
r}]+\frac{\delta (r)}{3k^{2}}I.
\end{eqnarray}%
In these equations, $k$ is the wave vector in the medium, $I$ is the three-order unit tensor, $r=|\mathbf{r}_{2}-\mathbf{r}_{1}|$ and $\hat{r}=%
\overrightarrow{r}/r$ is the unit vector. We denote the three-order tensor $%
\overleftrightarrow{G_{0}}$ with row and column indices $x,y,z$.

\subsection{Green Function of stratified medium\label{sec:stratifiedGF}}
Weyl Identity \cite{novotny2012principles} states that
\end{subequations}
\begin{equation}
\frac{\mathrm{e}^{\mathrm{i}kr}}{r}=\frac{\mathrm{i}}{2{\pi}}\int
\int \frac{\mathrm{e}^{\mathrm{i}(k_{x}x+k_{y}y+k_{z}|z|)}}{k_{z}}\mathrm{d}%
k_{x}\mathrm{d}k_{y}.
\end{equation}%
Using this identity we can decompose the free GF as planar wave with
different angular spectrum. Each planar wave is reflected by the interface
and all the reflected waves constitute the scattering GF. At the same time,
the scattering GF on the other side of the interface is constituted by
transmitted waves. The reflection and transmission coefficients are
described by Fresnel Law. Explicitly, the scattering GF by reflection is

\begin{eqnarray}
\overleftrightarrow{G_{s}}(\mathbf{r},\mathbf{r}_{0}) &=&\frac{\mathrm{i}}{8%
{\pi}^{2}}\iint_{-\infty }^{\infty }(\overleftrightarrow{M_{s}^{s}}+%
\overleftrightarrow{M_{s}^{p}})  \notag \\
&&\ast \mathrm{e}^{\mathrm{i}[k_{x}(x-x_{0})+k_{y}(y-y_{0})+k_{1z}(z+z_{1})]}%
\mathrm{d}k_{x}\mathrm{d}k_{y},
\end{eqnarray}%
where
\begin{subequations}
\begin{eqnarray}
\overleftrightarrow{M_{s}^{s}} &=&\frac{r^{s}(k_{x},k_{y})}{%
k_{1z}(k_{x}^{2}+k_{y}^{2})}%
\begin{bmatrix}
k_{y}^{2} & -k_{x}k_{y} & 0 \\
-k_{x}k_{y} & k_{x}^{2} & 0 \\
0 & 0 & 0%
\end{bmatrix}%
, \\
\overleftrightarrow{M_{s}^{p}} &=&\frac{-r^{p}(k_{x},k_{y})}{%
k_{1}^{2}(k_{x}^{2}+k_{y}^{2})}\ast  \notag \\
&&%
\begin{bmatrix}
k_{x}^{2}k_{1z} & k_{x}k_{y}k_{1z} & k_{x}(k_{x}^{2}+k_{y}^{2}) \\
k_{x}k_{y}k_{1z} & k_{y}^{2}k_{1z} & k_{y}(k_{x}^{2}+k_{y}^{2}) \\
-k_{x}(k_{x}^{2}+k_{y}^{2}) & -k_{y}(k_{x}^{2}+k_{y}^{2}) & -\frac{%
(k_{x}^{2}+k_{y}^{2})^{2}}{k_{1z}}%
\end{bmatrix}%
.
\end{eqnarray}%

In these equations, $r^{s}$ and $r^{p}$ are Fresnel reflection
coefficients and $k_{1z}$ is the $z$ component of wave vector in the incident
medium (denoted as medium $1$). Fresnel Law can be written in the wave vector
form:
\end{subequations}
\begin{subequations}
\begin{eqnarray}
r^{s}(k_{x},k_{y}) &=&\frac{\mu _{2}k_{1z}-\mu _{1}k_{2z}}{\mu
_{2}k_{1z}+\mu _{1}k_{2z}}, \\
r^{p}(k_{x},k_{y}) &=&\frac{\varepsilon _{2}k_{1z}-\varepsilon _{1}k_{2z}}{%
\varepsilon _{2}k_{1z}+\varepsilon _{1}k_{2z}}, \\
t^{s}(k_{x},k_{y}) &=&\frac{2\mu _{2}k_{1z}}{\mu _{2}k_{1z}+\mu _{1}k_{2z}},
\\
t^{p}(k_{x},k_{y}) &=&\frac{2\varepsilon _{2}k_{1z}}{\varepsilon
_{2}k_{1z}+\varepsilon _{1}k_{2z}}\sqrt{\frac{\mu _{2}\varepsilon _{1}}{\mu
_{1}\varepsilon _{2}}}.
\end{eqnarray}%
Fresnel Law of this form is valid for various kinds of wave vectors,
including those in cases of total reflection and metal reflection. For
single layer medium as we discuss here, Fresnel Law can be expressed as
\end{subequations}
\begin{subequations}
\begin{eqnarray}
r &=&\frac{r_{12}+r_{23}\mathrm{e}^{2\mathrm{i}k_{z2}d }}{1+r_{12}r_{23}%
\mathrm{e}^{2\mathrm{i}k_{z2}d }} \text{,}\\
t &=&\frac{t_{12}t_{23}\mathrm{e}^{\mathrm{i}k_{z2}d }}{1+r_{12}r_{23}%
\mathrm{e}^{2\mathrm{i}k_{z2}d }} \text{,}
\end{eqnarray}%
\end{subequations}
for $s$ and $p$ wave respectively, where $r_{12},r_{23},t_{12},t_{23}$ are
Fresnel coefficients on respective layer interfaces and $d$ is the thickness of the film. For multi-layer medium,
Fresnel coefficients have a similar form which can be got with recursion
\cite{chew1995waves}.

\bibliography{a-main.bbl}

\begin{thebibliography}{40}%
\makeatletter
\providecommand \@ifxundefined [1]{%
 \@ifx{#1\undefined}
}%
\providecommand \@ifnum [1]{%
 \ifnum #1\expandafter \@firstoftwo
 \else \expandafter \@secondoftwo
 \fi
}%
\providecommand \@ifx [1]{%
 \ifx #1\expandafter \@firstoftwo
 \else \expandafter \@secondoftwo
 \fi
}%
\providecommand \natexlab [1]{#1}%
\providecommand \enquote  [1]{``#1''}%
\providecommand \bibnamefont  [1]{#1}%
\providecommand \bibfnamefont [1]{#1}%
\providecommand \citenamefont [1]{#1}%
\providecommand \href@noop [0]{\@secondoftwo}%
\providecommand \href [0]{\begingroup \@sanitize@url \@href}%
\providecommand \@href[1]{\@@startlink{#1}\@@href}%
\providecommand \@@href[1]{\endgroup#1\@@endlink}%
\providecommand \@sanitize@url [0]{\catcode `\\12\catcode `\$12\catcode
  `\&12\catcode `\#12\catcode `\^12\catcode `\_12\catcode `\%12\relax}%
\providecommand \@@startlink[1]{}%
\providecommand \@@endlink[0]{}%
\providecommand \url  [0]{\begingroup\@sanitize@url \@url }%
\providecommand \@url [1]{\endgroup\@href {#1}{\urlprefix }}%
\providecommand \urlprefix  [0]{URL }%
\providecommand \Eprint [0]{\href }%
\providecommand \doibase [0]{https://doi.org/}%
\providecommand \selectlanguage [0]{\@gobble}%
\providecommand \bibinfo  [0]{\@secondoftwo}%
\providecommand \bibfield  [0]{\@secondoftwo}%
\providecommand \translation [1]{[#1]}%
\providecommand \BibitemOpen [0]{}%
\providecommand \bibitemStop [0]{}%
\providecommand \bibitemNoStop [0]{.\EOS\space}%
\providecommand \EOS [0]{\spacefactor3000\relax}%
\providecommand \BibitemShut  [1]{\csname bibitem#1\endcsname}%
\let\auto@bib@innerbib\@empty
\bibitem [{\citenamefont {Dicke}(1954)}]{DickePR1954}%
  \BibitemOpen
  \bibfield  {author} {\bibinfo {author} {\bibfnamefont {R.}~\bibnamefont
  {Dicke}},\ }\bibfield  {title} {\bibinfo {title} {Coherence in spontaneous
  radiation processes},\ }\href {https://doi.org/10.1103/PhysRev.93.99}
  {\bibfield  {journal} {\bibinfo  {journal} {Phys. Rev.}\ }\textbf {\bibinfo
  {volume} {93}},\ \bibinfo {pages} {99} (\bibinfo {year} {1954})}\BibitemShut
  {NoStop}%
\bibitem [{\citenamefont {Gross}\ and\ \citenamefont
  {Haroche}(1982)}]{GrossPR1982}%
  \BibitemOpen
  \bibfield  {author} {\bibinfo {author} {\bibfnamefont {M.}~\bibnamefont
  {Gross}}\ and\ \bibinfo {author} {\bibfnamefont {S.}~\bibnamefont
  {Haroche}},\ }\bibfield  {title} {\bibinfo {title} {Superradiance: An essay
  on the theory of collective spontaneous emission},\ }\href
  {https://doi.org/http://dx.doi.org/10.1016/0370-1573(82)90102-8} {\bibfield
  {journal} {\bibinfo  {journal} {Phys. Rep.}\ }\textbf {\bibinfo {volume}
  {93}},\ \bibinfo {pages} {301 } (\bibinfo {year} {1982})}\BibitemShut
  {NoStop}%
\bibitem [{\citenamefont {Scheibner}\ \emph {et~al.}(2007)\citenamefont
  {Scheibner}, \citenamefont {Schmidt}, \citenamefont {Worschech},
  \citenamefont {Forchel}, \citenamefont {Bacher}, \citenamefont {Passow},\
  and\ \citenamefont {Hommel}}]{scheibnerNPhysics2007superradiance}%
  \BibitemOpen
  \bibfield  {author} {\bibinfo {author} {\bibfnamefont {M.}~\bibnamefont
  {Scheibner}}, \bibinfo {author} {\bibfnamefont {T.}~\bibnamefont {Schmidt}},
  \bibinfo {author} {\bibfnamefont {L.}~\bibnamefont {Worschech}}, \bibinfo
  {author} {\bibfnamefont {A.}~\bibnamefont {Forchel}}, \bibinfo {author}
  {\bibfnamefont {G.}~\bibnamefont {Bacher}}, \bibinfo {author} {\bibfnamefont
  {T.}~\bibnamefont {Passow}},\ and\ \bibinfo {author} {\bibfnamefont
  {D.}~\bibnamefont {Hommel}},\ }\bibfield  {title} {\bibinfo {title}
  {Superradiance of quantum dots},\ }\href
  {http://www.nature.com/nphys/journal/v3/n2/abs/nphys494.html} {\bibfield
  {journal} {\bibinfo  {journal} {Nature Phys.}\ }\textbf {\bibinfo {volume}
  {3}},\ \bibinfo {pages} {106} (\bibinfo {year} {2007})}\BibitemShut {NoStop}%
\bibitem [{\citenamefont {Bernien}\ \emph {et~al.}(2013)\citenamefont
  {Bernien}, \citenamefont {Hensen}, \citenamefont {Pfaff}, \citenamefont
  {Koolstra}, \citenamefont {Blok}, \citenamefont {Robledo}, \citenamefont
  {Taminiau}, \citenamefont {Markham}, \citenamefont {Twitchen}, \citenamefont
  {Childress},\ and\ \citenamefont {Hanson}}]{BernienNature2013heralded}%
  \BibitemOpen
  \bibfield  {author} {\bibinfo {author} {\bibfnamefont {H.}~\bibnamefont
  {Bernien}}, \bibinfo {author} {\bibfnamefont {B.}~\bibnamefont {Hensen}},
  \bibinfo {author} {\bibfnamefont {W.}~\bibnamefont {Pfaff}}, \bibinfo
  {author} {\bibfnamefont {G.}~\bibnamefont {Koolstra}}, \bibinfo {author}
  {\bibfnamefont {M.}~\bibnamefont {Blok}}, \bibinfo {author} {\bibfnamefont
  {L.}~\bibnamefont {Robledo}}, \bibinfo {author} {\bibfnamefont
  {T.}~\bibnamefont {Taminiau}}, \bibinfo {author} {\bibfnamefont
  {M.}~\bibnamefont {Markham}}, \bibinfo {author} {\bibfnamefont
  {D.}~\bibnamefont {Twitchen}}, \bibinfo {author} {\bibfnamefont
  {L.}~\bibnamefont {Childress}},\ and\ \bibinfo {author} {\bibfnamefont
  {R.}~\bibnamefont {Hanson}},\ }\bibfield  {title} {\bibinfo {title} {Heralded
  entanglement between solid-state qubits separated by three metres},\ }\href
  {http://www.nature.com/nature/journal/v497/n7447/abs/nature12016.html}
  {\bibfield  {journal} {\bibinfo  {journal} {Nature}\ }\textbf {\bibinfo
  {volume} {497}},\ \bibinfo {pages} {86} (\bibinfo {year} {2013})}\BibitemShut
  {NoStop}%
\bibitem [{\citenamefont {Bouchet}\ \emph {et~al.}(2016)\citenamefont
  {Bouchet}, \citenamefont {Cao}, \citenamefont {Carminati}, \citenamefont
  {De~Wilde},\ and\ \citenamefont {Krachmalnicoff}}]{BouchetPRL2016Long}%
  \BibitemOpen
  \bibfield  {author} {\bibinfo {author} {\bibfnamefont {D.}~\bibnamefont
  {Bouchet}}, \bibinfo {author} {\bibfnamefont {D.}~\bibnamefont {Cao}},
  \bibinfo {author} {\bibfnamefont {R.}~\bibnamefont {Carminati}}, \bibinfo
  {author} {\bibfnamefont {Y.}~\bibnamefont {De~Wilde}},\ and\ \bibinfo
  {author} {\bibfnamefont {V.}~\bibnamefont {Krachmalnicoff}},\ }\bibfield
  {title} {\bibinfo {title} {Long-range plasmon-assisted energy transfer
  between fluorescent emitters},\ }\href
  {https://doi.org/10.1103/PhysRevLett.116.037401} {\bibfield  {journal}
  {\bibinfo  {journal} {Phys. Rev. Lett.}\ }\textbf {\bibinfo {volume} {116}},\
  \bibinfo {pages} {037401} (\bibinfo {year} {2016})}\BibitemShut {NoStop}%
\bibitem [{\citenamefont {Andrew}\ and\ \citenamefont
  {Barnes}(2004)}]{AndrewScience2004EnergyTansfer}%
  \BibitemOpen
  \bibfield  {author} {\bibinfo {author} {\bibfnamefont {P.}~\bibnamefont
  {Andrew}}\ and\ \bibinfo {author} {\bibfnamefont {W.~L.}\ \bibnamefont
  {Barnes}},\ }\bibfield  {title} {\bibinfo {title} {Energy transfer across a
  metal film mediated by surface plasmon polaritons},\ }\href
  {https://doi.org/10.1126/science.1102992} {\bibfield  {journal} {\bibinfo
  {journal} {Science}\ }\textbf {\bibinfo {volume} {306}},\ \bibinfo {pages}
  {1002} (\bibinfo {year} {2004})}\BibitemShut {NoStop}%
\bibitem [{\citenamefont {Dung}\ \emph {et~al.}(2002)\citenamefont {Dung},
  \citenamefont {Kn\"oll},\ and\ \citenamefont
  {Welsch}}]{DungPRA2002Intermolecular}%
  \BibitemOpen
  \bibfield  {author} {\bibinfo {author} {\bibfnamefont {H.~T.}\ \bibnamefont
  {Dung}}, \bibinfo {author} {\bibfnamefont {L.}~\bibnamefont {Kn\"oll}},\ and\
  \bibinfo {author} {\bibfnamefont {D.-G.}\ \bibnamefont {Welsch}},\ }\bibfield
   {title} {\bibinfo {title} {Intermolecular energy transfer in the presence of
  dispersing and absorbing media},\ }\href
  {https://doi.org/10.1103/PhysRevA.65.043813} {\bibfield  {journal} {\bibinfo
  {journal} {Phys. Rev. A}\ }\textbf {\bibinfo {volume} {65}},\ \bibinfo
  {pages} {043813} (\bibinfo {year} {2002})}\BibitemShut {NoStop}%
\bibitem [{\citenamefont {Marocico}\ and\ \citenamefont
  {Knoester}(2011)}]{MarocicoPRA2011Effect}%
  \BibitemOpen
  \bibfield  {author} {\bibinfo {author} {\bibfnamefont {C.~A.}\ \bibnamefont
  {Marocico}}\ and\ \bibinfo {author} {\bibfnamefont {J.}~\bibnamefont
  {Knoester}},\ }\bibfield  {title} {\bibinfo {title} {Effect of
  surface-plasmon polaritons on spontaneous emission and intermolecular
  energy-transfer rates in multilayered geometries},\ }\href
  {https://doi.org/10.1103/PhysRevA.84.053824} {\bibfield  {journal} {\bibinfo
  {journal} {Phys. Rev. A}\ }\textbf {\bibinfo {volume} {84}},\ \bibinfo
  {pages} {053824} (\bibinfo {year} {2011})}\BibitemShut {NoStop}%
\bibitem [{\citenamefont {Yang}\ \emph {et~al.}(2010)\citenamefont {Yang},
  \citenamefont {Xu}, \citenamefont {Song},\ and\ \citenamefont
  {Sun}}]{YangJCP2010}%
  \BibitemOpen
  \bibfield  {author} {\bibinfo {author} {\bibfnamefont {S.}~\bibnamefont
  {Yang}}, \bibinfo {author} {\bibfnamefont {D.~Z.}\ \bibnamefont {Xu}},
  \bibinfo {author} {\bibfnamefont {Z.}~\bibnamefont {Song}},\ and\ \bibinfo
  {author} {\bibfnamefont {C.~P.}\ \bibnamefont {Sun}},\ }\bibfield  {title}
  {\bibinfo {title} {Dimerization-assisted energy transport in light-harvesting
  complexes},\ }\href {https://doi.org/http://dx.doi.org/10.1063/1.3435213}
  {\bibfield  {journal} {\bibinfo  {journal} {J. Chem. Phys.}\ }\textbf
  {\bibinfo {volume} {132}},\ \bibinfo {eid} {234501} (\bibinfo {year}
  {2010})}\BibitemShut {NoStop}%
\bibitem [{\citenamefont {Benniston}\ and\ \citenamefont
  {Harriman}(2008)}]{BennistonMT2008}%
  \BibitemOpen
  \bibfield  {author} {\bibinfo {author} {\bibfnamefont {A.~C.}\ \bibnamefont
  {Benniston}}\ and\ \bibinfo {author} {\bibfnamefont {A.}~\bibnamefont
  {Harriman}},\ }\bibfield  {title} {\bibinfo {title} {Artificial
  photosynthesis},\ }\href
  {https://doi.org/http://dx.doi.org/10.1016/S1369-7021(08)70250-5} {\bibfield
  {journal} {\bibinfo  {journal} {Mater. Today}\ }\textbf {\bibinfo {volume}
  {11}},\ \bibinfo {pages} {26 } (\bibinfo {year} {2008})}\BibitemShut
  {NoStop}%
\bibitem [{\citenamefont {Wasielewski}(1992)}]{WasielewskiCR1992}%
  \BibitemOpen
  \bibfield  {author} {\bibinfo {author} {\bibfnamefont {M.~R.}\ \bibnamefont
  {Wasielewski}},\ }\bibfield  {title} {\bibinfo {title} {Photoinduced electron
  transfer in supramolecular systems for artificial photosynthesis},\ }\href
  {https://doi.org/10.1021/cr00011a005} {\bibfield  {journal} {\bibinfo
  {journal} {Chem. Rev.}\ }\textbf {\bibinfo {volume} {92}},\ \bibinfo {pages}
  {435} (\bibinfo {year} {1992})}\BibitemShut {NoStop}%
\bibitem [{\citenamefont {Sipahigil}\ \emph {et~al.}(2014)\citenamefont
  {Sipahigil}, \citenamefont {Jahnke}, \citenamefont {Rogers}, \citenamefont
  {Teraji}, \citenamefont {Isoya}, \citenamefont {Zibrov}, \citenamefont
  {Jelezko},\ and\ \citenamefont {Lukin}}]{SipahigilPRL2014}%
  \BibitemOpen
  \bibfield  {author} {\bibinfo {author} {\bibfnamefont {A.}~\bibnamefont
  {Sipahigil}}, \bibinfo {author} {\bibfnamefont {K.~D.}\ \bibnamefont
  {Jahnke}}, \bibinfo {author} {\bibfnamefont {L.~J.}\ \bibnamefont {Rogers}},
  \bibinfo {author} {\bibfnamefont {T.}~\bibnamefont {Teraji}}, \bibinfo
  {author} {\bibfnamefont {J.}~\bibnamefont {Isoya}}, \bibinfo {author}
  {\bibfnamefont {A.~S.}\ \bibnamefont {Zibrov}}, \bibinfo {author}
  {\bibfnamefont {F.}~\bibnamefont {Jelezko}},\ and\ \bibinfo {author}
  {\bibfnamefont {M.~D.}\ \bibnamefont {Lukin}},\ }\bibfield  {title} {\bibinfo
  {title} {Indistinguishable photons from separated silicon-vacancy centers in
  diamond},\ }\href {https://doi.org/10.1103/PhysRevLett.113.113602} {\bibfield
   {journal} {\bibinfo  {journal} {Phys. Rev. Lett.}\ }\textbf {\bibinfo
  {volume} {113}},\ \bibinfo {pages} {113602} (\bibinfo {year}
  {2014})}\BibitemShut {NoStop}%
\bibitem [{\citenamefont {Gruber}\ \emph {et~al.}(1997)\citenamefont {Gruber},
  \citenamefont {Dr\"{a}benstedt}, \citenamefont {Tietz}, \citenamefont
  {Fleury}, \citenamefont {Wrachtrup},\ and\ \citenamefont
  {Borczyskowski}}]{GruberScience1997}%
  \BibitemOpen
  \bibfield  {author} {\bibinfo {author} {\bibfnamefont {A.}~\bibnamefont
  {Gruber}}, \bibinfo {author} {\bibfnamefont {A.}~\bibnamefont
  {Dr\"{a}benstedt}}, \bibinfo {author} {\bibfnamefont {C.}~\bibnamefont
  {Tietz}}, \bibinfo {author} {\bibfnamefont {L.}~\bibnamefont {Fleury}},
  \bibinfo {author} {\bibfnamefont {J.}~\bibnamefont {Wrachtrup}},\ and\
  \bibinfo {author} {\bibfnamefont {C.~v.}\ \bibnamefont {Borczyskowski}},\
  }\bibfield  {title} {\bibinfo {title} {Scanning confocal optical microscopy
  and magnetic resonance on single defect centers},\ }\href
  {https://doi.org/10.1126/science.276.5321.2012} {\bibfield  {journal}
  {\bibinfo  {journal} {Science}\ }\textbf {\bibinfo {volume} {276}},\ \bibinfo
  {pages} {2012} (\bibinfo {year} {1997})}\BibitemShut {NoStop}%
\bibitem [{\citenamefont {Yamamoto}\ \emph {et~al.}(2013)\citenamefont
  {Yamamoto}, \citenamefont {M\"uller}, \citenamefont {McGuinness},
  \citenamefont {Teraji}, \citenamefont {Naydenov}, \citenamefont {Onoda},
  \citenamefont {Ohshima}, \citenamefont {Wrachtrup}, \citenamefont {Jelezko},\
  and\ \citenamefont {Isoya}}]{YamamotoPRB2013}%
  \BibitemOpen
  \bibfield  {author} {\bibinfo {author} {\bibfnamefont {T.}~\bibnamefont
  {Yamamoto}}, \bibinfo {author} {\bibfnamefont {C.}~\bibnamefont {M\"uller}},
  \bibinfo {author} {\bibfnamefont {L.}~\bibnamefont {McGuinness}}, \bibinfo
  {author} {\bibfnamefont {T.}~\bibnamefont {Teraji}}, \bibinfo {author}
  {\bibfnamefont {B.}~\bibnamefont {Naydenov}}, \bibinfo {author}
  {\bibfnamefont {S.}~\bibnamefont {Onoda}}, \bibinfo {author} {\bibfnamefont
  {T.}~\bibnamefont {Ohshima}}, \bibinfo {author} {\bibfnamefont
  {J.}~\bibnamefont {Wrachtrup}}, \bibinfo {author} {\bibfnamefont
  {F.}~\bibnamefont {Jelezko}},\ and\ \bibinfo {author} {\bibfnamefont
  {J.}~\bibnamefont {Isoya}},\ }\bibfield  {title} {\bibinfo {title} {Strongly
  coupled diamond spin qubits by molecular nitrogen implantation},\ }\href
  {https://doi.org/10.1103/PhysRevB.88.201201} {\bibfield  {journal} {\bibinfo
  {journal} {Phys. Rev. B}\ }\textbf {\bibinfo {volume} {88}},\ \bibinfo
  {pages} {201201} (\bibinfo {year} {2013})}\BibitemShut {NoStop}%
\bibitem [{\citenamefont {Jelezko}\ and\ \citenamefont
  {Wrachtrup}(2006)}]{JelezkoPSSA2006}%
  \BibitemOpen
  \bibfield  {author} {\bibinfo {author} {\bibfnamefont {F.}~\bibnamefont
  {Jelezko}}\ and\ \bibinfo {author} {\bibfnamefont {J.}~\bibnamefont
  {Wrachtrup}},\ }\bibfield  {title} {\bibinfo {title} {Single defect centres
  in diamond: A review},\ }\href {https://doi.org/10.1002/pssa.200671403}
  {\bibfield  {journal} {\bibinfo  {journal} {Phys. Stat. Sol. (a)}\ }\textbf
  {\bibinfo {volume} {203}},\ \bibinfo {pages} {3207} (\bibinfo {year}
  {2006})}\BibitemShut {NoStop}%
\bibitem [{\citenamefont {Doherty}\ \emph {et~al.}(2013)\citenamefont
  {Doherty}, \citenamefont {Manson}, \citenamefont {Delaney}, \citenamefont
  {Jelezko}, \citenamefont {Wrachtrup},\ and\ \citenamefont
  {Hollenberg}}]{DohertyPR2014}%
  \BibitemOpen
  \bibfield  {author} {\bibinfo {author} {\bibfnamefont {M.~W.}\ \bibnamefont
  {Doherty}}, \bibinfo {author} {\bibfnamefont {N.~B.}\ \bibnamefont {Manson}},
  \bibinfo {author} {\bibfnamefont {P.}~\bibnamefont {Delaney}}, \bibinfo
  {author} {\bibfnamefont {F.}~\bibnamefont {Jelezko}}, \bibinfo {author}
  {\bibfnamefont {J.}~\bibnamefont {Wrachtrup}},\ and\ \bibinfo {author}
  {\bibfnamefont {L.~C.}\ \bibnamefont {Hollenberg}},\ }\bibfield  {title}
  {\bibinfo {title} {The nitrogen-vacancy colour centre in diamond},\ }\href
  {https://doi.org/http://dx.doi.org/10.1016/j.physrep.2013.02.001} {\bibfield
  {journal} {\bibinfo  {journal} {Phys. Rep.}\ }\textbf {\bibinfo {volume}
  {528}},\ \bibinfo {pages} {1 } (\bibinfo {year} {2013})}\BibitemShut
  {NoStop}%
\bibitem [{\citenamefont {Rittweger}\ \emph {et~al.}(2009)\citenamefont
  {Rittweger}, \citenamefont {Han}, \citenamefont {Irvine}, \citenamefont
  {Eggeling},\ and\ \citenamefont {Hell}}]{Rittweger2009sted}%
  \BibitemOpen
  \bibfield  {author} {\bibinfo {author} {\bibfnamefont {E.}~\bibnamefont
  {Rittweger}}, \bibinfo {author} {\bibfnamefont {K.~Y.}\ \bibnamefont {Han}},
  \bibinfo {author} {\bibfnamefont {S.~E.}\ \bibnamefont {Irvine}}, \bibinfo
  {author} {\bibfnamefont {C.}~\bibnamefont {Eggeling}},\ and\ \bibinfo
  {author} {\bibfnamefont {S.~W.}\ \bibnamefont {Hell}},\ }\bibfield  {title}
  {\bibinfo {title} {Sted microscopy reveals crystal colour centres with
  nanometric resolution},\ }\href
  {http://www.nature.com/nphoton//journal/v3/n3/full/nphoton.2009.2.html}
  {\bibfield  {journal} {\bibinfo  {journal} {Nature Photon.}\ }\textbf
  {\bibinfo {volume} {3}},\ \bibinfo {pages} {144} (\bibinfo {year}
  {2009})}\BibitemShut {NoStop}%
\bibitem [{\citenamefont {Cui}\ \emph {et~al.}(2013)\citenamefont {Cui},
  \citenamefont {Sun}, \citenamefont {Chen}, \citenamefont {Gong},\ and\
  \citenamefont {Guo}}]{Cui2013PRL}%
  \BibitemOpen
  \bibfield  {author} {\bibinfo {author} {\bibfnamefont {J.-M.}\ \bibnamefont
  {Cui}}, \bibinfo {author} {\bibfnamefont {F.-W.}\ \bibnamefont {Sun}},
  \bibinfo {author} {\bibfnamefont {X.-D.}\ \bibnamefont {Chen}}, \bibinfo
  {author} {\bibfnamefont {Z.-J.}\ \bibnamefont {Gong}},\ and\ \bibinfo
  {author} {\bibfnamefont {G.-C.}\ \bibnamefont {Guo}},\ }\bibfield  {title}
  {\bibinfo {title} {Quantum statistical imaging of particles without
  restriction of the diffraction limit},\ }\href
  {https://doi.org/10.1103/PhysRevLett.110.153901} {\bibfield  {journal}
  {\bibinfo  {journal} {Phys. Rev. Lett.}\ }\textbf {\bibinfo {volume} {110}},\
  \bibinfo {pages} {153901} (\bibinfo {year} {2013})}\BibitemShut {NoStop}%
\bibitem [{\citenamefont {Chen}\ \emph {et~al.}(2015)\citenamefont {Chen},
  \citenamefont {Zou}, \citenamefont {Gong}, \citenamefont {Dong},
  \citenamefont {Guo},\ and\ \citenamefont {Sun}}]{Chen2014LSA}%
  \BibitemOpen
  \bibfield  {author} {\bibinfo {author} {\bibfnamefont {X.}~\bibnamefont
  {Chen}}, \bibinfo {author} {\bibfnamefont {C.}~\bibnamefont {Zou}}, \bibinfo
  {author} {\bibfnamefont {Z.}~\bibnamefont {Gong}}, \bibinfo {author}
  {\bibfnamefont {C.}~\bibnamefont {Dong}}, \bibinfo {author} {\bibfnamefont
  {G.}~\bibnamefont {Guo}},\ and\ \bibinfo {author} {\bibfnamefont
  {F.}~\bibnamefont {Sun}},\ }\bibfield  {title} {\bibinfo {title}
  {{Subdiffraction optical manipulation of the charge state of nitrogen vacancy
  center in diamond}},\ }\href {https://doi.org/10.1038/lsa.2015.3} {\bibfield
  {journal} {\bibinfo  {journal} {Light: Sci. Appl.}\ }\textbf {\bibinfo
  {volume} {4}},\ \bibinfo {pages} {e230} (\bibinfo {year} {2015})}\BibitemShut
  {NoStop}%
\bibitem [{\citenamefont {Pfender}\ \emph {et~al.}(2014)\citenamefont
  {Pfender}, \citenamefont {Aslam}, \citenamefont {Waldherr}, \citenamefont
  {Neumann},\ and\ \citenamefont {Wrachtrup}}]{PfenderPNAS2014SingleSpin}%
  \BibitemOpen
  \bibfield  {author} {\bibinfo {author} {\bibfnamefont {M.}~\bibnamefont
  {Pfender}}, \bibinfo {author} {\bibfnamefont {N.}~\bibnamefont {Aslam}},
  \bibinfo {author} {\bibfnamefont {G.}~\bibnamefont {Waldherr}}, \bibinfo
  {author} {\bibfnamefont {P.}~\bibnamefont {Neumann}},\ and\ \bibinfo {author}
  {\bibfnamefont {J.}~\bibnamefont {Wrachtrup}},\ }\bibfield  {title} {\bibinfo
  {title} {{Single-spin stochastic optical reconstruction microscopy}},\ }\href
  {https://doi.org/10.1073/pnas.1404907111} {\bibfield  {journal} {\bibinfo
  {journal} {Proc. Natl. Acad. Sci.}\ }\textbf {\bibinfo {volume} {111}},\
  \bibinfo {pages} {14669} (\bibinfo {year} {2014})}\BibitemShut {NoStop}%
\bibitem [{\citenamefont {Ford}\ and\ \citenamefont
  {Weber}(1984)}]{ford1984electromagnetic}%
  \BibitemOpen
  \bibfield  {author} {\bibinfo {author} {\bibfnamefont {G.~W.}\ \bibnamefont
  {Ford}}\ and\ \bibinfo {author} {\bibfnamefont {W.}~\bibnamefont {Weber}},\
  }\bibfield  {title} {\bibinfo {title} {Electromagnetic interactions of
  molecules with metal surfaces},\ }\href
  {http://www.sciencedirect.com/science/article/pii/037015738490098X}
  {\bibfield  {journal} {\bibinfo  {journal} {Phys. Rep.}\ }\textbf {\bibinfo
  {volume} {113}},\ \bibinfo {pages} {195} (\bibinfo {year}
  {1984})}\BibitemShut {NoStop}%
\bibitem [{\citenamefont {Gonzalez-Tudela}\ \emph {et~al.}(2011)\citenamefont
  {Gonzalez-Tudela}, \citenamefont {Martin-Cano}, \citenamefont {Moreno},
  \citenamefont {Martin-Moreno}, \citenamefont {Tejedor},\ and\ \citenamefont
  {Garcia-Vidal}}]{GonzalezPRLentanglement}%
  \BibitemOpen
  \bibfield  {author} {\bibinfo {author} {\bibfnamefont {A.}~\bibnamefont
  {Gonzalez-Tudela}}, \bibinfo {author} {\bibfnamefont {D.}~\bibnamefont
  {Martin-Cano}}, \bibinfo {author} {\bibfnamefont {E.}~\bibnamefont {Moreno}},
  \bibinfo {author} {\bibfnamefont {L.}~\bibnamefont {Martin-Moreno}}, \bibinfo
  {author} {\bibfnamefont {C.}~\bibnamefont {Tejedor}},\ and\ \bibinfo {author}
  {\bibfnamefont {F.~J.}\ \bibnamefont {Garcia-Vidal}},\ }\bibfield  {title}
  {\bibinfo {title} {Entanglement of two qubits mediated by one-dimensional
  plasmonic waveguides},\ }\href
  {https://doi.org/10.1103/PhysRevLett.106.020501} {\bibfield  {journal}
  {\bibinfo  {journal} {Phys. Rev. Lett.}\ }\textbf {\bibinfo {volume} {106}},\
  \bibinfo {pages} {020501} (\bibinfo {year} {2011})}\BibitemShut {NoStop}%
\bibitem [{\citenamefont {Zhou}\ \emph {et~al.}(2011)\citenamefont {Zhou},
  \citenamefont {Liu},\ and\ \citenamefont {Li}}]{ZhouOL2011Optics}%
  \BibitemOpen
  \bibfield  {author} {\bibinfo {author} {\bibfnamefont {F.}~\bibnamefont
  {Zhou}}, \bibinfo {author} {\bibfnamefont {Y.}~\bibnamefont {Liu}},\ and\
  \bibinfo {author} {\bibfnamefont {Z.-Y.}\ \bibnamefont {Li}},\ }\bibfield
  {title} {\bibinfo {title} {Surface-plasmon-polariton-assisted dipole--dipole
  interaction near metal surfaces},\ }\href
  {https://doi.org/10.1364/OL.36.001969} {\bibfield  {journal} {\bibinfo
  {journal} {Opt. Lett.}\ }\textbf {\bibinfo {volume} {36}},\ \bibinfo {pages}
  {1969} (\bibinfo {year} {2011})}\BibitemShut {NoStop}%
\bibitem [{\citenamefont {Dzsotjan}\ \emph {et~al.}(2011)\citenamefont
  {Dzsotjan}, \citenamefont {K\"astel},\ and\ \citenamefont
  {Fleischhauer}}]{Dzsotjan2011PRB}%
  \BibitemOpen
  \bibfield  {author} {\bibinfo {author} {\bibfnamefont {D.}~\bibnamefont
  {Dzsotjan}}, \bibinfo {author} {\bibfnamefont {J.}~\bibnamefont {K\"astel}},\
  and\ \bibinfo {author} {\bibfnamefont {M.}~\bibnamefont {Fleischhauer}},\
  }\bibfield  {title} {\bibinfo {title} {Dipole-dipole shift of quantum
  emitters coupled to surface plasmons of a nanowire},\ }\href
  {https://doi.org/10.1103/PhysRevB.84.075419} {\bibfield  {journal} {\bibinfo
  {journal} {Phys. Rev. B}\ }\textbf {\bibinfo {volume} {84}},\ \bibinfo
  {pages} {075419} (\bibinfo {year} {2011})}\BibitemShut {NoStop}%
\bibitem [{\citenamefont {Meijer}\ \emph {et~al.}(2005)\citenamefont {Meijer},
  \citenamefont {Burchard}, \citenamefont {Domhan}, \citenamefont {Wittmann},
  \citenamefont {Gaebel}, \citenamefont {Popa}, \citenamefont {Jelezko},\ and\
  \citenamefont {Wrachtrup}}]{MeijerAPL2005Generation}%
  \BibitemOpen
  \bibfield  {author} {\bibinfo {author} {\bibfnamefont {J.}~\bibnamefont
  {Meijer}}, \bibinfo {author} {\bibfnamefont {B.}~\bibnamefont {Burchard}},
  \bibinfo {author} {\bibfnamefont {M.}~\bibnamefont {Domhan}}, \bibinfo
  {author} {\bibfnamefont {C.}~\bibnamefont {Wittmann}}, \bibinfo {author}
  {\bibfnamefont {T.}~\bibnamefont {Gaebel}}, \bibinfo {author} {\bibfnamefont
  {I.}~\bibnamefont {Popa}}, \bibinfo {author} {\bibfnamefont {F.}~\bibnamefont
  {Jelezko}},\ and\ \bibinfo {author} {\bibfnamefont {J.}~\bibnamefont
  {Wrachtrup}},\ }\bibfield  {title} {\bibinfo {title} {Generation of single
  color centers by focused nitrogen implantation},\ }\href
  {http://scitation.aip.org/content/aip/journal/apl/87/26/10.1063/1.2103389}
  {\bibfield  {journal} {\bibinfo  {journal} {Appl. Phys. Lett.}\ }\textbf
  {\bibinfo {volume} {87}},\ \bibinfo {eid} {261909} (\bibinfo {year}
  {2005})}\BibitemShut {NoStop}%
\bibitem [{\citenamefont {Wang}\ \emph {et~al.}(2015)\citenamefont {Wang},
  \citenamefont {Feng}, \citenamefont {Zhang}, \citenamefont {Chen},
  \citenamefont {Zheng}, \citenamefont {Guo}, \citenamefont {Zhang},
  \citenamefont {Song}, \citenamefont {Guo}, \citenamefont {Fan}, \citenamefont
  {Zou}, \citenamefont {Lou}, \citenamefont {Zhu},\ and\ \citenamefont
  {Wang}}]{WangPRB2015High-sensitivity}%
  \BibitemOpen
  \bibfield  {author} {\bibinfo {author} {\bibfnamefont {J.}~\bibnamefont
  {Wang}}, \bibinfo {author} {\bibfnamefont {F.}~\bibnamefont {Feng}}, \bibinfo
  {author} {\bibfnamefont {J.}~\bibnamefont {Zhang}}, \bibinfo {author}
  {\bibfnamefont {J.}~\bibnamefont {Chen}}, \bibinfo {author} {\bibfnamefont
  {Z.}~\bibnamefont {Zheng}}, \bibinfo {author} {\bibfnamefont
  {L.}~\bibnamefont {Guo}}, \bibinfo {author} {\bibfnamefont {W.}~\bibnamefont
  {Zhang}}, \bibinfo {author} {\bibfnamefont {X.}~\bibnamefont {Song}},
  \bibinfo {author} {\bibfnamefont {G.}~\bibnamefont {Guo}}, \bibinfo {author}
  {\bibfnamefont {L.}~\bibnamefont {Fan}}, \bibinfo {author} {\bibfnamefont
  {C.}~\bibnamefont {Zou}}, \bibinfo {author} {\bibfnamefont {L.}~\bibnamefont
  {Lou}}, \bibinfo {author} {\bibfnamefont {W.}~\bibnamefont {Zhu}},\ and\
  \bibinfo {author} {\bibfnamefont {G.}~\bibnamefont {Wang}},\ }\bibfield
  {title} {\bibinfo {title} {High-sensitivity temperature sensing using an
  implanted single nitrogen-vacancy center array in diamond},\ }\href
  {https://doi.org/10.1103/PhysRevB.91.155404} {\bibfield  {journal} {\bibinfo
  {journal} {Phys. Rev. B}\ }\textbf {\bibinfo {volume} {91}},\ \bibinfo
  {pages} {155404} (\bibinfo {year} {2015})}\BibitemShut {NoStop}%
\bibitem [{\citenamefont {Ohno}\ \emph {et~al.}(2012)\citenamefont {Ohno},
  \citenamefont {Joseph~Heremans}, \citenamefont {Bassett}, \citenamefont
  {Myers}, \citenamefont {Toyli}, \citenamefont {Bleszynski~Jayich},
  \citenamefont {Palmstrom},\ and\ \citenamefont
  {Awschalom}}]{OhnoAPL2012Engineering}%
  \BibitemOpen
  \bibfield  {author} {\bibinfo {author} {\bibfnamefont {K.}~\bibnamefont
  {Ohno}}, \bibinfo {author} {\bibfnamefont {F.}~\bibnamefont
  {Joseph~Heremans}}, \bibinfo {author} {\bibfnamefont {L.~C.}\ \bibnamefont
  {Bassett}}, \bibinfo {author} {\bibfnamefont {B.~A.}\ \bibnamefont {Myers}},
  \bibinfo {author} {\bibfnamefont {D.~M.}\ \bibnamefont {Toyli}}, \bibinfo
  {author} {\bibfnamefont {A.~C.}\ \bibnamefont {Bleszynski~Jayich}}, \bibinfo
  {author} {\bibfnamefont {C.~J.}\ \bibnamefont {Palmstrom}},\ and\ \bibinfo
  {author} {\bibfnamefont {D.~D.}\ \bibnamefont {Awschalom}},\ }\bibfield
  {title} {\bibinfo {title} {Engineering shallow spins in diamond with nitrogen
  delta-doping},\ }\href
  {http://scitation.aip.org/content/aip/journal/apl/101/8/10.1063/1.4748280}
  {\bibfield  {journal} {\bibinfo  {journal} {Appl. Phys. Lett.}\ }\textbf
  {\bibinfo {volume} {101}},\ \bibinfo {eid} {082413} (\bibinfo {year}
  {2012})}\BibitemShut {NoStop}%
\bibitem [{\citenamefont {Wang}\ \emph {et~al.}(2016)\citenamefont {Wang},
  \citenamefont {Zhang}, \citenamefont {Zhang}, \citenamefont {You},
  \citenamefont {Li}, \citenamefont {Guo}, \citenamefont {Feng}, \citenamefont
  {Song}, \citenamefont {Lou}, \citenamefont {Zhu},\ and\ \citenamefont
  {Wang}}]{WangNanoscale2016coherence}%
  \BibitemOpen
  \bibfield  {author} {\bibinfo {author} {\bibfnamefont {J.}~\bibnamefont
  {Wang}}, \bibinfo {author} {\bibfnamefont {W.}~\bibnamefont {Zhang}},
  \bibinfo {author} {\bibfnamefont {J.}~\bibnamefont {Zhang}}, \bibinfo
  {author} {\bibfnamefont {J.}~\bibnamefont {You}}, \bibinfo {author}
  {\bibfnamefont {Y.}~\bibnamefont {Li}}, \bibinfo {author} {\bibfnamefont
  {G.}~\bibnamefont {Guo}}, \bibinfo {author} {\bibfnamefont {F.}~\bibnamefont
  {Feng}}, \bibinfo {author} {\bibfnamefont {X.}~\bibnamefont {Song}}, \bibinfo
  {author} {\bibfnamefont {L.}~\bibnamefont {Lou}}, \bibinfo {author}
  {\bibfnamefont {W.}~\bibnamefont {Zhu}},\ and\ \bibinfo {author}
  {\bibfnamefont {G.}~\bibnamefont {Wang}},\ }\bibfield  {title} {\bibinfo
  {title} {Coherence times of precise depth controlled nv centers in diamond},\
  }\href {https://doi.org/10.1039/C5NR08690F} {\bibfield  {journal} {\bibinfo
  {journal} {Nanoscale}\ }\textbf {\bibinfo {volume} {8}},\ \bibinfo {pages}
  {5780} (\bibinfo {year} {2016})}\BibitemShut {NoStop}%
\bibitem [{\citenamefont {McLellan}\ \emph {et~al.}(2016)\citenamefont
  {McLellan}, \citenamefont {Myers}, \citenamefont {Kraemer}, \citenamefont
  {Ohno}, \citenamefont {Awschalom},\ and\ \citenamefont
  {Jayich}}]{ClaireNL2016patterned}%
  \BibitemOpen
  \bibfield  {author} {\bibinfo {author} {\bibfnamefont {C.~A.}\ \bibnamefont
  {McLellan}}, \bibinfo {author} {\bibfnamefont {B.~A.}\ \bibnamefont {Myers}},
  \bibinfo {author} {\bibfnamefont {S.}~\bibnamefont {Kraemer}}, \bibinfo
  {author} {\bibfnamefont {K.}~\bibnamefont {Ohno}}, \bibinfo {author}
  {\bibfnamefont {D.~D.}\ \bibnamefont {Awschalom}},\ and\ \bibinfo {author}
  {\bibfnamefont {A.~C.~B.}\ \bibnamefont {Jayich}},\ }\bibfield  {title}
  {\bibinfo {title} {Patterned formation of highly coherent nitrogen-vacancy
  centers using a focused electron irradiation technique},\ }\href
  {https://doi.org/10.1021/acs.nanolett.5b05304} {\bibfield  {journal}
  {\bibinfo  {journal} {Nano Lett.}\ }\textbf {\bibinfo {volume} {16}},\
  \bibinfo {pages} {2450} (\bibinfo {year} {2016})}\BibitemShut {NoStop}%
\bibitem [{\citenamefont {Maze}\ \emph {et~al.}(2011)\citenamefont {Maze},
  \citenamefont {Gali}, \citenamefont {Togan}, \citenamefont {Chu},
  \citenamefont {Trifonov}, \citenamefont {Kaxiras},\ and\ \citenamefont
  {Lukin}}]{MazeNJP2011}%
  \BibitemOpen
  \bibfield  {author} {\bibinfo {author} {\bibfnamefont {J.~R.}\ \bibnamefont
  {Maze}}, \bibinfo {author} {\bibfnamefont {A.}~\bibnamefont {Gali}}, \bibinfo
  {author} {\bibfnamefont {E.}~\bibnamefont {Togan}}, \bibinfo {author}
  {\bibfnamefont {Y.}~\bibnamefont {Chu}}, \bibinfo {author} {\bibfnamefont
  {A.}~\bibnamefont {Trifonov}}, \bibinfo {author} {\bibfnamefont
  {E.}~\bibnamefont {Kaxiras}},\ and\ \bibinfo {author} {\bibfnamefont {M.~D.}\
  \bibnamefont {Lukin}},\ }\bibfield  {title} {\bibinfo {title} {Properties of
  nitrogen-vacancy centers in diamond: the group theoretic approach},\ }\href
  {http://stacks.iop.org/1367-2630/13/i=2/a=025025} {\bibfield  {journal}
  {\bibinfo  {journal} {New J. Phys.}\ }\textbf {\bibinfo {volume} {13}},\
  \bibinfo {pages} {025025} (\bibinfo {year} {2011})}\BibitemShut {NoStop}%
\bibitem [{\citenamefont {Manson}\ \emph {et~al.}(2006)\citenamefont {Manson},
  \citenamefont {Harrison},\ and\ \citenamefont {Sellars}}]{MansonPRB2006}%
  \BibitemOpen
  \bibfield  {author} {\bibinfo {author} {\bibfnamefont {N.~B.}\ \bibnamefont
  {Manson}}, \bibinfo {author} {\bibfnamefont {J.~P.}\ \bibnamefont
  {Harrison}},\ and\ \bibinfo {author} {\bibfnamefont {M.~J.}\ \bibnamefont
  {Sellars}},\ }\bibfield  {title} {\bibinfo {title} {Nitrogen-vacancy center
  in diamond: Model of the electronic structure and associated dynamics},\
  }\href {https://doi.org/10.1103/PhysRevB.74.104303} {\bibfield  {journal}
  {\bibinfo  {journal} {Phys. Rev. B}\ }\textbf {\bibinfo {volume} {74}},\
  \bibinfo {pages} {104303} (\bibinfo {year} {2006})}\BibitemShut {NoStop}%
\bibitem [{\citenamefont {Paulus}\ \emph {et~al.}(2000)\citenamefont {Paulus},
  \citenamefont {Gay-Balmaz},\ and\ \citenamefont {Martin}}]{PaulusPRE2000}%
  \BibitemOpen
  \bibfield  {author} {\bibinfo {author} {\bibfnamefont {M.}~\bibnamefont
  {Paulus}}, \bibinfo {author} {\bibfnamefont {P.}~\bibnamefont {Gay-Balmaz}},\
  and\ \bibinfo {author} {\bibfnamefont {O.~J.~F.}\ \bibnamefont {Martin}},\
  }\bibfield  {title} {\bibinfo {title} {Accurate and efficient computation of
  the green's tensor for stratified media},\ }\href
  {https://doi.org/10.1103/PhysRevE.62.5797} {\bibfield  {journal} {\bibinfo
  {journal} {Phys. Rev. E}\ }\textbf {\bibinfo {volume} {62}},\ \bibinfo
  {pages} {5797} (\bibinfo {year} {2000})}\BibitemShut {NoStop}%
\bibitem [{\citenamefont {Novotny}\ and\ \citenamefont
  {Hecht}(2012)}]{novotny2012principles}%
  \BibitemOpen
  \bibfield  {author} {\bibinfo {author} {\bibfnamefont {L.}~\bibnamefont
  {Novotny}}\ and\ \bibinfo {author} {\bibfnamefont {B.}~\bibnamefont
  {Hecht}},\ }\href
  {https://books.google.com/books?hl=zh-CN&lr=&id=RHC_AwAAQBAJ&oi=fnd&pg=PR15&dq=Principles+of+nano-optics&ots=XZVCLIuiEI&sig=mRgj544ceWcrHUk8YQtjFOLRF_Q}
  {\emph {\bibinfo {title} {Principles of nano-optics}}}\ (\bibinfo
  {publisher} {Cambridge university press},\ \bibinfo {year}
  {2012})\BibitemShut {NoStop}%
\bibitem [{\citenamefont {Ruppin}\ and\ \citenamefont
  {Martin}(2004)}]{RuppinJCP2004}%
  \BibitemOpen
  \bibfield  {author} {\bibinfo {author} {\bibfnamefont {R.}~\bibnamefont
  {Ruppin}}\ and\ \bibinfo {author} {\bibfnamefont {O.~J.~F.}\ \bibnamefont
  {Martin}},\ }\bibfield  {title} {\bibinfo {title} {Lifetime of an emitting
  dipole near various types of interfaces including magnetic and negative
  refractive materials},\ }\href
  {http://scitation.aip.org/content/aip/journal/jcp/121/22/10.1063/1.1812742}
  {\bibfield  {journal} {\bibinfo  {journal} {J. Chem. Phys.}\ }\textbf
  {\bibinfo {volume} {121}},\ \bibinfo {pages} {11358} (\bibinfo {year}
  {2004})}\BibitemShut {NoStop}%
\bibitem [{\citenamefont {Polyanskiy}()}]{PolyanskiyWebBookRefractive}%
  \BibitemOpen
  \bibfield  {author} {\bibinfo {author} {\bibfnamefont {M.~N.}\ \bibnamefont
  {Polyanskiy}},\ }\href {http://refractiveindex.info} {\emph {\bibinfo {title}
  {Refractive index database}}}\ (\bibinfo  {publisher} {Online; accessed
  2-June-2016})\BibitemShut {NoStop}%
\bibitem [{\citenamefont {Garc\'{i}a~de Abajo}(2007)}]{AbajoRMP2007}%
  \BibitemOpen
  \bibfield  {author} {\bibinfo {author} {\bibfnamefont {F.~J.}\ \bibnamefont
  {Garc\'{i}a~de Abajo}},\ }\bibfield  {title} {\bibinfo {title} {Colloquium:
  Light scattering by particle and hole arrays},\ }\href
  {https://doi.org/10.1103/RevModPhys.79.1267} {\bibfield  {journal} {\bibinfo
  {journal} {Rev. Mod. Phys.}\ }\textbf {\bibinfo {volume} {79}},\ \bibinfo
  {pages} {1267} (\bibinfo {year} {2007})}\BibitemShut {NoStop}%
\bibitem [{\citenamefont {Rotenberg}\ \emph {et~al.}(2012)\citenamefont
  {Rotenberg}, \citenamefont {Spasenovi\'{c}}, \citenamefont {Krijger},
  \citenamefont {le~Feber}, \citenamefont {Garc\'{i}a~de Abajo},\ and\
  \citenamefont {Kuipers}}]{RotenbergPRL2012}%
  \BibitemOpen
  \bibfield  {author} {\bibinfo {author} {\bibfnamefont {N.}~\bibnamefont
  {Rotenberg}}, \bibinfo {author} {\bibfnamefont {M.}~\bibnamefont
  {Spasenovi\'{c}}}, \bibinfo {author} {\bibfnamefont {T.~L.}\ \bibnamefont
  {Krijger}}, \bibinfo {author} {\bibfnamefont {B.}~\bibnamefont {le~Feber}},
  \bibinfo {author} {\bibfnamefont {F.~J.}\ \bibnamefont {Garc\'{i}a~de
  Abajo}},\ and\ \bibinfo {author} {\bibfnamefont {L.}~\bibnamefont
  {Kuipers}},\ }\bibfield  {title} {\bibinfo {title} {Plasmon scattering from
  single subwavelength holes},\ }\href
  {https://doi.org/10.1103/PhysRevLett.108.127402} {\bibfield  {journal}
  {\bibinfo  {journal} {Phys. Rev. Lett.}\ }\textbf {\bibinfo {volume} {108}},\
  \bibinfo {pages} {127402} (\bibinfo {year} {2012})}\BibitemShut {NoStop}%
\bibitem [{\citenamefont {Burke}\ \emph {et~al.}(1986)\citenamefont {Burke},
  \citenamefont {Stegeman},\ and\ \citenamefont {Tamir}}]{BurkePRB1986surface}%
  \BibitemOpen
  \bibfield  {author} {\bibinfo {author} {\bibfnamefont {J.~J.}\ \bibnamefont
  {Burke}}, \bibinfo {author} {\bibfnamefont {G.~I.}\ \bibnamefont
  {Stegeman}},\ and\ \bibinfo {author} {\bibfnamefont {T.}~\bibnamefont
  {Tamir}},\ }\bibfield  {title} {\bibinfo {title} {Surface-polariton-like
  waves guided by thin, lossy metal films},\ }\href
  {https://doi.org/10.1103/PhysRevB.33.5186} {\bibfield  {journal} {\bibinfo
  {journal} {Phys. Rev. B}\ }\textbf {\bibinfo {volume} {33}},\ \bibinfo
  {pages} {5186} (\bibinfo {year} {1986})}\BibitemShut {NoStop}%
\bibitem [{\citenamefont {Mart{\'\i}n-Cano}\ \emph {et~al.}(2010)\citenamefont
  {Mart{\'\i}n-Cano}, \citenamefont {Martín-Moreno}, \citenamefont
  {García-Vidal},\ and\ \citenamefont {Moreno}}]{MartinNL2010ResonanceEnergy}%
  \BibitemOpen
  \bibfield  {author} {\bibinfo {author} {\bibfnamefont {D.}~\bibnamefont
  {Mart{\'\i}n-Cano}}, \bibinfo {author} {\bibfnamefont {L.}~\bibnamefont
  {Martín-Moreno}}, \bibinfo {author} {\bibfnamefont {F.~J.}\ \bibnamefont
  {García-Vidal}},\ and\ \bibinfo {author} {\bibfnamefont {E.}~\bibnamefont
  {Moreno}},\ }\bibfield  {title} {\bibinfo {title} {Resonance energy transfer
  and superradiance mediated by plasmonic nanowaveguides},\ }\href
  {http://pubs.acs.org/doi/abs/10.1021/nl101876f} {\bibfield  {journal}
  {\bibinfo  {journal} {Nano Lett.}\ }\textbf {\bibinfo {volume} {10}},\
  \bibinfo {pages} {3129} (\bibinfo {year} {2010})}\BibitemShut {NoStop}%
\bibitem [{\citenamefont {Chew}(1995)}]{chew1995waves}%
  \BibitemOpen
  \bibfield  {author} {\bibinfo {author} {\bibfnamefont {W.~C.}\ \bibnamefont
  {Chew}},\ }\href
  {http://as.wiley.com/WileyCDA/WileyTitle/productCd-0780347498,miniSiteCd-IEEE2.html}
  {\emph {\bibinfo {title} {Waves and fields in inhomogeneous media}}},\ Vol.\
  \bibinfo {volume} {522}\ (\bibinfo  {publisher} {IEEE press New York},\
  \bibinfo {year} {1995})\BibitemShut {NoStop}%
\end{thebibliography}%

\end{document}